\documentclass[epj]{svjour}   
\usepackage{mathptmx} % use Times fonts if available on your TeX system
\usepackage[colorlinks,citecolor=blue,urlcolor=blue,linkcolor=blue]{hyperref}
\usepackage{amsmath,amssymb,cite,graphicx,bigstrut}
\usepackage{slashed}
\usepackage{bm}
\usepackage[caption=false]{subfig}
\usepackage{mathtools}
\usepackage{refcount}
\usepackage[usenames,dvipsnames,svgnames,table]{xcolor}
\newcommand{\imag}{\mathrm{Im}}
\renewcommand{\Im}{\mathrm{Im}\,} 

\newcommand{\nn}{\nonumber}
\newcommand{\mk}{m_{K}}
\newcommand{\mpi}{m_{\pi}}
\def\hhref#1{\href{http://arxiv.org/abs/#1}{arXiv:#1}} % href in bibliography

\journalname{EPJ}
\begin{document}
\flushbottom
\sloppy

\title{
Dispersive treatment of $K_S\to\gamma\gamma$ and $K_S\to\gamma\ell^+\ell^-$}

\author{
Gilberto Colangelo
\and
Ramon Stucki
\and
Lewis~C.~Tunstall
}

\institute{
Albert Einstein Center for Fundamental Physics, Institute for Theoretical
Physics,  \\ 
University of Bern, Sidlerstrasse 5, 3012 Bern, Switzerland \label{addr1}
}

\date{September 2016} 

\abstract{We analyse the rare kaon decays $K_S \to \gamma\gamma$ and 
$K_S \to \gamma\ell^+\ell^-$ $(\ell = e \mbox{ or } \mu)$ in a dispersive 
framework in which the weak Hamiltonian carries momentum. Our analysis extends 
predictions from lowest-order $SU(3)_L\times SU(3)_R$ chiral perturbation theory 
($\chi$PT$_3$) to fully account for effects from final-state interactions, and 
is free from ambiguities associated with extrapolating the kaon off-shell.  
Given input from $K_S \to \pi\pi$ and $\gamma\gamma^{(*)}\to\pi\pi$, we solve 
the once-subtracted dispersion relations numerically to predict the rates for 
$K_S \to \gamma\gamma$ and $K_S \to \gamma\ell^+\ell^-$.  In the leptonic modes, 
we find sizeable corrections to the $\chi$PT$_3$ predictions for the integrated 
rates.
\PACS{
      {13.20.Eb}{Decays of $K$ mesons}   \and
      {11.55.Fv}{Dispersion relations}
      }
}

\maketitle

%%%%%%%%%%%%%%%%%%%%%%%%
\section{Introduction}%%
\label{sec:intro}%%%%%%%
%%%%%%%%%%%%%%%%%%%%%%%%
In the study of kaon decays, our ability to obtain precise predictions from the 
Standard Model (SM) depends on whether the underlying physics is predominantly 
of short- or long-distance nature.  At one end of a broad spectrum of possible 
decay channels, there are ``golden modes'' like $K\to \pi \nu\bar\nu$, where the 
amplitude factorises into a hadronic form factor and perturbative corrections 
--- both of which are under excellent theoretical control~\cite{Buras:2015qea}.  
In such cases, the resulting  prediction can be at a level of precision that 
competes with (or even surpasses) current experimental measurements. This state 
of affairs can lead to powerful constraints on physics beyond the SM and drives 
much of the theoretical and experimental interest in these modes.

By contrast, non-leptonic decays such as $K\to \pi\pi$ and $K\to\pi\pi\pi$ are 
dominated by long-distance contributions involving hadronic matrix elements of 
four-quark operators.  The evaluation of these matrix elements is a notoriously 
difficult non-perturbative problem, and this hinders the comparison of theory 
with experiment.  

In between these extremes lies a range of decay modes in which 
a clean separation of the short- and long-distance physics can be achieved with 
varying degrees of success.

Since kaon decays occur at low energies, a systematic analysis can be undertaken 
within $SU(3)_L\times SU(3)_R$ chiral perturbation theory ($\chi$PT$_3$), where 
amplitudes are expanded as an asymptotic series in powers of $O(\mk)$ momentum 
and light quark masses $m_{u,d,s} = O(\mk^2)$.  The application of $\chi$PT$_3$ 
to kaon decays is covered in a comprehensive review~\cite{Cirigliano:2011ny}; 
here we recall two important features that determine the quality of predictions 
arising from the 3-flavour expansion:
\begin{enumerate}
  \item hadronic uncertainties are parametrised in terms of low-energy 
  constants (LECs), whose values are not fixed by chiral symmetry alone.  For 
  several purely leptonic and semi-leptonic kaon decays, the corresponding 
  LECs can be extracted from a combination of experimental data and input from 
  lattice QCD.  However, the situation for non-leptonic and weak radiative 
  decays is far less certain, with many of the LECs essentially unconstrained at 
  next-to-lowest-order (NLO) in the chiral expansion; 
  \item at energies above the $\pi\pi$ threshold, final-state interactions 
  (FSI), especially in the $0^{++}$ 
  channel~\cite{Neveu:1970tn,Truong:1984uu,Truong:1987hn,Dobado:1989qm}, can 
  spoil the convergence of the $\chi$PT$_3$ expansion.  These effects are 
  related to the broad $f_0(500)$ resonance~\cite{Pelaez:2015qba}, whose 
  $O(\mk)$ mass~\cite{Caprini:2005zr} implies a lack of scale separation between 
  the Goldstone $\pi,K,\eta$ and non-Goldstone $f_0,\rho,\omega,\ldots$ sectors. 
  In these cases, chiral-perturbative methods must be abandoned in favour of 
  non-perturbative methods based on unitarity, analyticity, and crossing 
  symmetry.%
  \footnote{Scale separation can be restored in scenarios where $f_0$ belongs to 
  the Goldstone sector, as in chiral-scale perturbation 
  theory~\cite{Crewther:2012wd,Crewther:2013vea}.}
\end{enumerate}

Dispersion relations offer a means to address items 1 and 2 within a 
model-independent framework.  These methods have been mostly applied in the 
context of pure strong processes such as pion form
factors~\cite{Donoghue:1990xh,Ananthanarayan:2004xy}, 
$\pi\pi$-scattering~\cite{Roy:1971tc,Ananthanarayan:2000ht,Colangelo:2001df,DescotesGenon:2001tn,Kaminski:2006qe},
$\pi K$-scattering~\cite{Buettiker:2003pp}, 
$\gamma \gamma^{(*)} \to \pi\pi$ 
\cite{GarciaMartin:2010cw,Hoferichter:2011wk,Moussallam:2013una}, $\pi N$ 
scattering~\cite{Ditsche:2012fv,Hoferichter:2012wf,Hoferichter:2015hva}, 
semi-leptonic kaon decays $K_{\ell3}$~\cite{Jamin:2001zq,Jamin:2004re,Jamin:2006tj,Bernard:2006gy,Bernard:2009zm,Abbas:2010ns} 
and $K_{\ell4}$~\cite{Truong:1980pf,Bijnens:1994ie,Colangelo:2015kha}, or decays 
not involving kaons, e.g.\ $\eta \to \pi\pi\pi$ 
\cite{Roiesnel:1980gd,Kambor:1995yc,Anisovich:1996tx,Colangelo:2011zz,Kampf:2011wr,Guo:2015zqa}.  

In view of current high-statistics kaon experiments such as NA62 
\cite{Collazuol:2009zz}, we believe it is timely to consider extending the scope 
of dispersive methods to $\Delta S=1$ processes involving the effective weak 
Hamiltonian ${\cal H}_w$, and in particular to two-body decays. Such an 
extension was proposed sometime ago by B\"uchler \emph{et al.} 
\cite{Buchler:2001nm,Colangelo:2001uv}, who treated the decay $K\to\pi\pi$ 
dispersively by allowing ${\cal H}_w$ to carry momentum, thereby overcoming the 
difficulty that the kinematics in two-body decays are completely fixed. The 
advantage of this approach over $\chi$PT$_3$ is that (a) only a few subtraction 
constants are required as input, and (b) $\pi\pi$ rescattering effects are fully 
accounted for in terms of Omn\`{e}s factors and calculable dispersive integrals 
in crossed channels.  Moreover, by allowing ${\cal H}_w$ to carry momentum, the 
ambiguities associated with taking the kaon off-shell 
\cite{Colangelo:2001uv,Buchler:2001np} are entirely avoided.

In this article, we extend the dispersive framework developed 
in~\cite{Buchler:2001nm} to the rare decays $K_S \to \gamma\gamma$ and 
$K_S\to \gamma\ell^+\ell^-$ $(\ell = e \mbox{ or } \mu)$.  In lowest-order (LO) 
$\chi$PT$_3$, the amplitudes for $K_S\to \gamma\gamma^{(*)}$ possess the well 
known feature of ultraviolet finite $\pi^\pm$, $K^\pm$ one-loop diagrams coupled 
to the external photons.  For the pure radiative decay, the chiral 
prediction~\cite{D'Ambrosio:1986ze,Goity:1986sr,Cirigliano:2011ny} for the rate 
\begin{equation}
  \mathrm{BR}(K_S \to \gamma\gamma)_{\chi\mathrm{PT}_3} = 2.0 \times 10^{-6}
\end{equation}
is in reasonable agreement with the experimental average~\cite{Agashe:2014kda}
\begin{equation}
  \mathrm{BR}(K_S \to \gamma\gamma) = (2.63\pm 0.17)\times 10^{-6}\,,
  \label{BR Kgg}
\end{equation}
while the predictions~\cite{Ecker:1987hd} for the leptonic modes are typically 
expressed in terms of the ratios
\begin{equation}
  \left.\frac{\Gamma(K_S\to\gamma\ell^+\ell^-)}{\Gamma(K_S\to\gamma\gamma)}
  \right|_{\chi\mathrm{PT}_3} 
  = \left\{ 
  \begin{array}{cl} 
  1.6\times 10^{-2} & \quad(\ell = e) \\ 
  3.8\times 10^{-4} & \quad(\ell=\mu)
  \end{array} \right.\,.
  \label{chpt Kgll}
\end{equation}
Although these decays have not yet been measured, they may lie within reach of 
the KLOE-2 experiment at DA$\mathrm{\Phi}$NE~\cite{AmelinoCamelia:2010me}, which 
is projected to be sensitive down to $K_S$ branching ratios of $O(10^{-9})$.  
Given these projections, it is clearly of interest to determine what impact 
$\pi\pi$ rescattering effects have on the $\chi$PT$_3$ predictions 
(\ref{chpt Kgll}). 

The outline of this paper is as follows.  In Section~\ref{sec:prelim} we 
introduce the general formalism needed to analyse $K_S\to\gamma\gamma^*$  
dispersively, and derive the decomposition of the decay amplitude into a basis 
of scalar functions that are free from kinematic zeros and singularities.  
In particular, we use this basis to extend the LO $\chi$PT$_3$ 
calculation~\cite{Ecker:1987hd} to the case where ${\cal H}_w$ carries non-zero 
momentum.  Section~\ref{sec:Ktopipi} reviews the dispersive framework developed 
for $K_S\to\pi\pi$~\cite{Buchler:2001nm}, which forms a key input 
in our analysis of $K_S\to\gamma\gamma^*$. In Section~\ref{sec:Ktogg} we examine 
$K_S\to\gamma\gamma$ and find that the inclusion of effects from FSI improves 
the agreement between theory and experiment.  We also comment on how our results 
compare with previous work~\cite{Kambor:1993tv} based on extrapolating the kaon 
off-shell.  Section~\ref{sec:Ktogll} concerns $K_S\to\gamma\ell^+\ell^-$, where 
we observe that FSI and the pion vector form factor lead to sizeable corrections 
of the LO $\chi$PT$_3$ predictions.  Our summary is given in 
Section~\ref{sec:summary}.

%%%%%%%%%%%%%%%%%%%%%%%%%
\section{Preliminaries}%%
\label{sec:prelim}%%%%%%%
%%%%%%%%%%%%%%%%%%%%%%%%%
We begin by considering the radiative decay
\begin{equation}
  K_S(k) \to \gamma(q_1)\gamma^*(q_2)\,,
\end{equation}
whose amplitude is given by
\begin{align}
M(K_S\to\gamma\gamma^*) = 
e^2\epsilon^{\mu*}_1(q_1,\lambda_1)\epsilon^{\nu*}_2(q_2,\lambda_2)
A_{\mu\nu}(k,q_1,q_2)\,,
\label{KK matrix}
\end{align}
where $\epsilon_{1,2}$ are the polarization vectors of the photons.  The tensor 
$A_{\mu\nu}$ is defined in terms of the pure $\Delta I =1/2$ matrix element%
  \footnote{In non-leptonic $\Delta S=1$ processes, it is observed that 
  amplitudes with $\Delta I =1/2$ dominate over other isospin transitions.  As 
  in~\cite{Buchler:2001nm}, we focus on this dominant contribution to 
  $K_S \to \gamma\gamma^*$, noting that the dispersive framework can easily be 
  adapted to a determination of the sub-dominant $\Delta I=3/2$ amplitude. \label{rule}}
\begin{align}
  &A_{\mu\nu}(k,q_1,q_2) = \\
  &- \int\!\! d^4x\, d^4y\, e^{i(q_1 \cdot x + q_2 \cdot y)} 
  \langle\mathrm{vac} | T\{ J_\mu(x)J_\nu(y) {\cal H}_w^{1/2}(0)\}|K_S(k)\rangle 
  \,, \nn
\end{align}
where $J_\mu$ is the electromagnetic current of the light quarks $u,d,s$, and we 
allow the weak Hamiltonian ${\cal H}_w$ to carry non-zero momentum 
$h_\mu \neq 0$.  Then the decay amplitude (\ref{KK matrix}) becomes a function 
of the three Mandelstam variables
\begin{equation}
  s = (q_1 + q_2)^2 \,, \quad t = (k-q_1)^2\,, \quad u = (k-q_2)^2\,,
\end{equation}
which satisfy
\begin{equation}
  s+t+u = \mk^2 + q_2^2 + h^2\,.
\end{equation}
In what follows it is convenient to set $h^2=0$, while keeping $h_\mu \neq 0$ in 
general.  Doing so does not result in a loss of generality, but does 
simplify several expressions derived in this paper. To recover the physical 
decay amplitude, one simply takes the limit $h_\mu \to 0$, in which case the 
kinematic variables become fixed at the values
\begin{equation}
s = \mk^2\,, \quad t=q_2^2\,, \quad u=0\,.
\label{phys pt}
\end{equation} 

%%%%%%%%%%%%%%%%%%%%%%%%%%%%%%%%%%%
\subsection{Tensor decomposition}%%
%%%%%%%%%%%%%%%%%%%%%%%%%%%%%%%%%%%
To set up a dispersive framework for $K_S\to\gamma\gamma^*$, the first step is 
to decompose $A_{\mu\nu}$ in a basis of independent tensors, whose scalar 
coefficients are free from kinematic singularities and zeros.  This can be 
achieved by applying the prescription of Bardeen and Tung~\cite{Bardeen:1969aw}, 
and Tarrach~\cite{Tarrach:1975tu}; our approach resembles the tensor 
decomposition of $\gamma^*\gamma^*\to\pi\pi$ discussed 
in~\cite{Colangelo:2014dfa,Stoffer:2014rka,Colangelo:2015ama}.

Let $q_i = \{q_1,\,q_2,\,h-k\}$ label the three independent momenta and observe 
that Lorentz covariance and $CP$-invariance implies a decomposition in 
terms of ten tensors%
  \footnote{The terms 
  $\sim \sum_{i,j}\epsilon_{\mu\nu\rho\sigma} q_i^\rho q_j^\sigma A_3^{ij}$ 
  are allowed by Lorentz covariance, but violate $P$  and $CP$ symmetry.}
\begin{equation}
  A_{\mu\nu} = g_{\mu\nu} A_1 + \sum_{i,j=1}^3 q_{i\mu}q_{j\nu} A_2^{ij}\,.
  \label{Amunu gen}
\end{equation}
The scalar functions $\{A_1,\, A_2^{ij}\}$ are not all independent since 
$A_{\mu\nu}$ is constrained by the electromagnetic Ward identities
\begin{equation}
  q_1^\mu A_{\mu\nu} = q_2^\nu A_{\mu\nu} = 0\,.
  \label{WIA}
\end{equation}
A convenient way to impose the constraint (\ref{WIA}) is to introduce the 
gauge projector
\begin{equation}
  P_{\mu\nu} = g_{\mu\nu} - \frac{q_{2\mu}q_{1\nu}}{q_1\cdot q_2}\,,
\end{equation}
and let it act on both indices of $A_{\mu\nu}$:
\begin{equation}
  A_{\mu\nu} = P_{\mu\alpha} P_{\beta\nu} A^{\alpha\beta} 
  = \sum_{i=1}^5 \bar T_{\mu\nu}^i \bar A_i\,.
\end{equation}
By definition, this leaves the physical tensor $A_{\mu\nu}$ invariant and 
removes contributions that do not satisfy the Ward identities; with this 
procedure the set of scalar functions reduces to five.  The new basis functions 
$\bar A_i$ are free from kinematic singularities, but contain zeros because the 
tensors $\bar T_{\mu\nu}^i$ contain single and double poles in $q_1\cdot q_2$.  
As shown in~\cite{Colangelo:2014dfa,Stoffer:2014rka,Colangelo:2015ama}, the 
removal of these poles can be performed by adding suitable linear combinations 
of $\bar T_{\mu\nu}^i$ with non-singular coefficients, followed by a rescaling 
in powers of $q_1\cdot q_2$.  In our case, contraction with $\epsilon_1$ and 
setting $q_1^2=0$ imposes two additional constraints, so the final result is
\begin{equation}
  A_{\mu\nu}(k,q_1,q_2) = \sum_{i=1}^3 T_{\mu\nu}^i\, B_i(s,t,u,q_2^2)\,,
  \label{Amunu decomp}
\end{equation}
where the scalar functions $B_i$ are free from kinematic zeros and 
singularities, and the corresponding tensors are
\begin{align}
    T^1_{\mu\nu}  &= (q_1 \cdot q_2) g_{\mu\nu} - q_{2\mu} q_{1\nu} \,, \\
    T^2_{\mu\nu}  &= (q_1 \cdot q_2) q_{3\mu} q_{2\nu} 
                   - q_2^2 q_{3\mu} q_{1\nu} \nn \\
                  &+\tfrac{1}{2}\big[(t-u)-\mk^2\big]
                   (q_2^2 g_{\mu\nu}-q_{2\mu} q_{2\nu})\,,\nn \\
    T^3_{\mu\nu}  &= (q_1 \cdot q_2) q_{3\mu} q_{3\nu} 
                   - \tfrac{1}{4} \big[ (t-u)^2 - \mk^4 \big] g_{\mu\nu} \nn \\
                  &+ \tfrac{1}{2} \big[(t-u) +\mk^2 \big] q_{3\mu} q_{1\nu} 
                   - \tfrac{1}{2} \big[(t-u) -\mk^2 \big]q_{2\mu} q_{3\nu}\,.\nn
\end{align}
At the physical point (\ref{phys pt}) there are only two independent momenta, so 
$A_{\mu\nu}$ reduces to $T_{\mu\nu}^1$ times the coefficient
\begin{equation}
  B_1(\mk^2,q_2^2) - q_2^2\, B_2(\mk^2,q_2^2) 
  + \tfrac{1}{2}(q_2^2+\mk^2)B_3(\mk^2,q_2^2)\,.
\end{equation}
Evidently, the determination of the scalar functions $B_i$ completely fixes the 
prediction for the $K_S \to \gamma\gamma^*$ amplitude (\ref{KK matrix}).

%%%%%%%%%%%%%%%%%%%%%%%%%%%%%%%%%%%%%%%%%%%%%%%%%%%%%%%%%%%%%%%%%%%%%%%%%%%%%%
\subsection{$K_S\to\gamma\gamma^{*}$ in lowest order $\chi$PT$_{\mathbf 3}$}%%
\label{sec:Kggstar ChPT}
%%%%%%%%%%%%%%%%%%%%%%%%%%%%%%%%%%%%%%%%%%%%%%%%%%%%%%%%%%%%%%%%%%%%%%%%%%%%%%
\begin{figure}[t]
  \centering\includegraphics[scale=0.41]{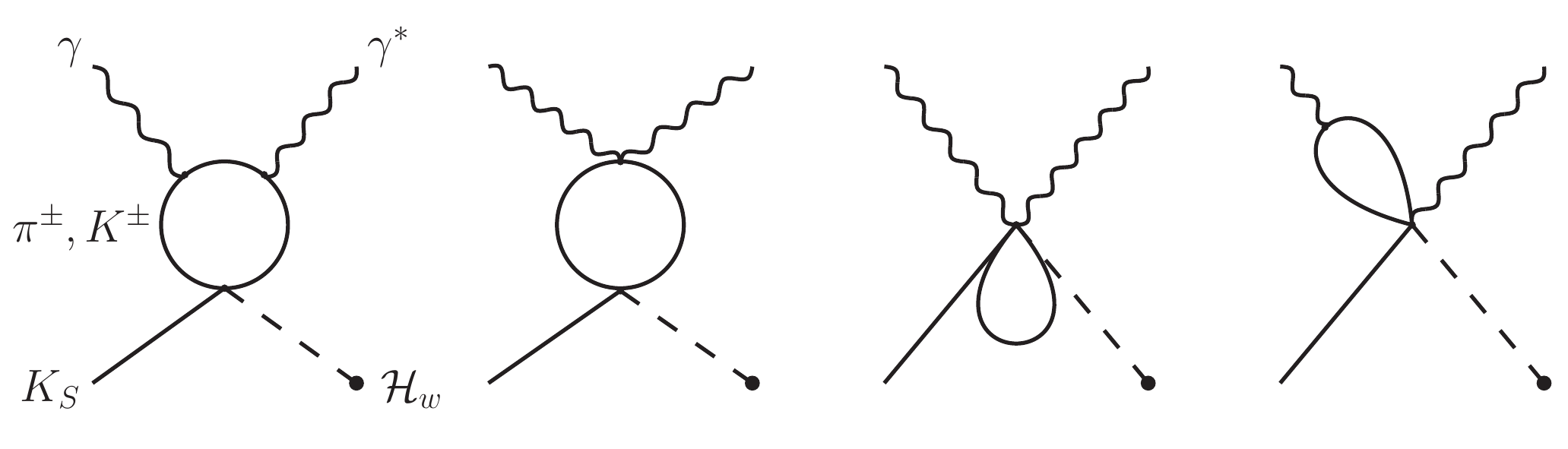}
  \caption{Lowest order $\chi$PT$_3$ graphs for $K_S\to\gamma\gamma^*$, where 
  the weak Hamiltonian ${\cal H}_w$ carries momentum.}
  \label{fig:kk_2gam}
\end{figure}
Before discussing our dispersive treatment of the scalar functions $B_i$, it is 
instructive to extend the LO $\chi$PT$_3$ calculation of 
$K_S\to\gamma\gamma^*$~\cite{Ecker:1987hd} to the case where ${\cal H}_w$ 
carries momentum. In the conventions of~\cite{Cirigliano:2011ny}, the graphs 
shown in Figure \ref{fig:kk_2gam} yield
\begin{equation}
  A_{\mu\nu}\big|_{{\chi\mathrm{PT}_3}} = -iG_8 F_\pi 
  ( 3s +\mk^2 -4\mpi^2) I_{\mu\nu} -\big\{\mpi^2 \to \mk^2\big\}\,, 
\end{equation}
where $G_8 = 9.1 \times 10^{-6}$ GeV$^{-2}$ is the octet coupling at $O(p^2)$,
$F_\pi = 92.2$ MeV is the pion decay constant~\cite{Agashe:2014kda}, and the 
loop integral is
\begin{equation}
  I_{\mu\nu} = \int\!\!\! \frac{d^{4}\ell}{(2\pi)^4} 
  \frac{g_{\mu\nu}(\ell^2-m_\phi^2) - (2\ell+q_1)_\mu(2\ell-q_2)_\nu}
  {[(\ell+q_1)^2-m_\phi^2][(\ell-q_2)^2-m_\phi^2][\ell^2-m_\phi^2]}\,, 
\end{equation}
where $\phi = \pi^\pm$ or $K^\pm$. The integral is ultraviolet finite and 
can be evaluated in terms of Feynman parameters:
\begin{equation}
  I_{\mu\nu} = \frac{i}{16\pi^2}\int_0^1 du \int_0^{1-u} dv 
  \frac{4uv\, T_{\mu\nu}^1 - 2v(1-2v) T_{\mu\nu}^4}{D(u,v,m_\phi^2)}\,,
  \label{Imunu}
\end{equation}
where the denominator is given by
\begin{equation}
  D(u,v,m_\phi^2) = m_\phi^2 - suv - v(1-u-v) q_2^2 - i\epsilon \,.
\end{equation}
In (\ref{Imunu}), the second tensor
\begin{equation}
  T_{\mu\nu}^4 = (q_1\cdot q_2) q_{1\mu}q_{2\nu} - q_2^2 q_{1\mu}q_{1\nu}
\end{equation}
vanishes upon contraction with $\epsilon_1$, so we find that only $B_1$ 
contributes to $M(K_S\to\gamma\gamma^*)$ at LO, with
\begin{align}
  &B_{1}(s,q_2^2)\big|_{\chi\mathrm{PT}_3}   \label{Aggstar chpt}  \\
  &=\frac{G_8 F_\pi }{4\pi^2}
  \bigg( \frac{3s +\mk^2 -4\mpi^2}{s}\bigg) H(s,\mpi^2,q_2^2) 
  - \big\{\mpi^2\to\mk^2\big\}\,. \nn
  \end{align}
Here, the quantity
\begin{align}
  &H(s,m^2,q^2) = \frac{s^2}{2(s-q^2)^2} \\
  &\times \bigg\{ \frac{q^2}{s} F\bigg(\frac{q^2}{m^2}\bigg) 
  - F\bigg(\frac{s}{m^2}\bigg) -\frac{2q^2}{s} 
  \bigg[ G\bigg(\frac{q^2}{m^2}\bigg) - G\bigg(\frac{s}{m^2}\bigg) 
  \bigg] \bigg\} \nn
\end{align}
is defined~\cite{Cirigliano:2011ny} in terms of the one-loop functions
\begin{align}
  F(a) &= \left\{ \begin{array}{lll} 
  1 - \dfrac{4}{a}\arcsin^2\big(\sqrt{a}/2\big) & \quad & a \leq 4 \,, \\
  1 + \dfrac{1}{a}\bigg(\ln \dfrac{1-\sqrt{1-4/a}}{1+\sqrt{1-4/a}}+i\pi\bigg)^2 & \quad & a > 4\,,
  \end{array}\right.  \\
  G(a) &= \left\{ \def\arraystretch{2}\begin{array}{lll} 
  \sqrt{4/a-1}\arcsin\big(\sqrt{a}/2\big) & \quad & a \leq 4\,, \\
  \dfrac{1}{2}\sqrt{1-4/a}\bigg(\ln \dfrac{1+\sqrt{1-4/a}}{1-\sqrt{1-4/a}} - i\pi \bigg) & \quad & a>4\,.
  \end{array}\right. \nn
\end{align}

At the physical point (\ref{phys pt}), the expression in (\ref{Aggstar chpt}) 
agrees with the original $\chi$PT$_3$ result~\cite{Ecker:1987hd}, as it should. 

As emphasised in~\cite{Crewther:1985zt}, tadpole cancellation 
completely eliminates the weak mass operator at $O(p^2)$ in the $\chi$PT$_3$ 
expansion. The argument can be extended to $O(p^4)$ \cite{Kambor:1989tz} and 
remains valid when ${\cal H}_w$ carries momentum.

%%%%%%%%%%%%%%%%%%%%%%%%%%%%%%%%%%%%%%%%%%%%%%%%%%%%%%%%%
\subsection{Unitarity and $\pi\pi$ intermediate states}%%
%%%%%%%%%%%%%%%%%%%%%%%%%%%%%%%%%%%%%%%%%%%%%%%%%%%%%%%%%
\begin{figure}[t]
  \centering\includegraphics[scale=0.8]{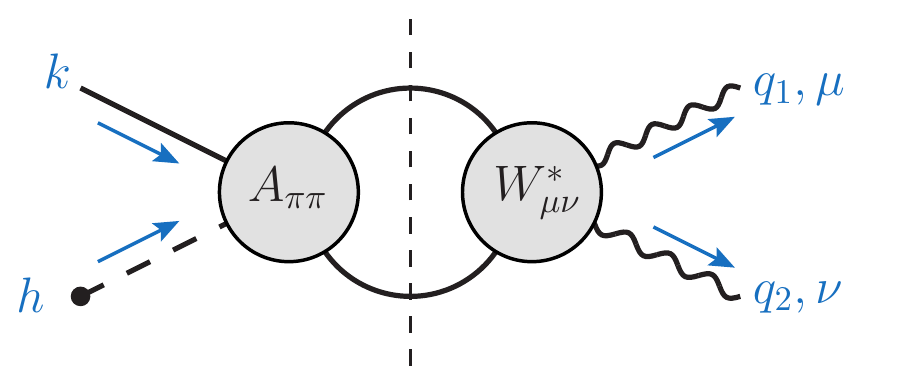}
  \caption{Unitarity relation for the $\pi\pi$ intermediate state in 
  $K_S\to\gamma\gamma^*$, where the weak Hamiltonian carries momentum 
  $h_\mu \neq 0$. The dashed line indicates the cutting of the pion propagators, 
  while the grey blobs refer to the respective $K_S\to\pi\pi$ and 
  $\gamma\gamma^*\to\pi\pi$ sub-amplitudes.}
  \label{fig:kk_gg_cut}
\end{figure}
Let us now analyse the unitarity relation due to the intermediate $\pi\pi$ 
state.  In the $s$-channel, this contribution reads (Figure \ref{fig:kk_gg_cut})
\begin{align}
  \mathrm{disc}_s\, A_{\mu\nu} &= \frac{1}{2} \int\!\!\! 
  \frac{d^{3}p_1}{(2\pi)^32E_1} \frac{d^{3}p_2}{(2\pi)^32E_2}(2\pi)^4 
  \label{phase} \\
  &\times \delta^4(q_1+q_2- p_1 - p_2)
   A_{\pi\pi}(s,t',u')W_{\mu\nu}^*(q_1,q_2,p_1)\,,  \nn
\end{align}
where $A_{\pi\pi}$ and $W_{\mu\nu}$ are the amplitudes for the subprocesses 
$K_S\to\pi\pi$ and $\gamma\gamma^*\to\pi\pi$ respectively.  On the left-hand 
side of the cut, the Mandelstam variables are
\begin{equation}
  t' = (k-p_1)^2 \,, \quad u' = (k-p_2)^2\,,
  \label{lhs cut}
\end{equation}
while on the right-hand side, $W_{\mu\nu}$ can be decomposed into a basis of  
three independent tensors~\cite{Colangelo:2014dfa,Stoffer:2014rka,Colangelo:2015ama,Moussallam:2013una}:
\begin{equation}
  W_{\mu\nu}(q_1,q_2,p_1) = \sum_{i=1}^3 t_{\mu\nu}^i W_i(s,t'',u'',q_2^2)\,,
\end{equation}
where 
\begin{equation}
  t''= (q_1 - p_1)^2 \,, \qquad u'' = (q_1 - p_2)^2\,,
\end{equation}
and 
\begin{align}
  t_{\mu\nu}^1 &= (q_1\cdot q_2)g_{\mu\nu} - q_{2\mu}q_{1\nu}\,, \nn \\
  t_{\mu\nu}^2 &= (q_1\cdot q_2)\Delta_\mu (q_{2\nu} - q_2^2 q_{1\nu}) 
  + \tfrac{1}{2}(t''-u'')(q_2^2 g_{\mu\nu} - q_{2\mu}q_{2\nu})\,, \nn \\
  t_{\mu\nu}^3 &= (q_1\cdot q_2) \Delta_\mu \Delta_\nu 
                - \tfrac{1}{4}(t''-u'')^2g_{\mu\nu} \nn \\
  &+ \tfrac{1}{2}(t''-u'')(\Delta_\mu q_{1\nu} - q_{2\mu}\Delta_\nu) \,, 
  \qquad \Delta = p_2 - p_1\,.
\end{align}

The phase space integration (\ref{phase}) must project each of the tensors 
$t_{\mu\nu}^i$ onto linear combinations of $T_{\mu\nu}^{i}$.  However, the 
integration is trivial if contributions from $D$ waves and higher are neglected. 
This is because in this approximation, $A_{\pi\pi}$ is independent of $t'$ and 
$u'$, while the scalar functions $W_i$ can be expressed in terms of a single 
helicity partial wave~\cite{Stoffer:2014rka,Colangelo:2015ama},
\begin{align}
  W_1 &= -\frac{2}{s-q_2^2}h_{++}^0(s,q_2^2)\,, \quad W_2 = W_3 = 0\,.
  \label{W swave}
\end{align}
Since $W_{1}$ is independent of the pion momenta, the tensor $t_{\mu\nu}^{1}$ 
can be pulled under the phase space integral (\ref{phase}).  Equating the scalar 
coefficients then gives the analytical result
\begin{align}
  \mathrm{disc}_s\, B_1(s,q_2^2) &= \notag \\
  &\frac{1}{32\pi^2} \int_0^\infty\!\!\! dp 
  \frac{p^2}{\mpi^2+p^2} \delta\bigg(q_1^0+q_2^0 - 2\sqrt{\mpi^2+p^2}\bigg)
  \notag \\
  &\times 
   \int \!\! d\Omega'' A_{\pi\pi}(s) W_1^*(s,q_2^2) \notag \\
  &= -\frac{\sigma_\pi(s)}{8\pi}\, 
  \frac{A_{\pi\pi}(s)[h_{++}^0(s,q_2^2)]^*}{s-q_2^2}\,,
  \label{ImA12}
\end{align}
where we have introduced the kinematic factor 
\begin{equation}
  \sigma_\phi(s) = \sqrt{1-4m_\phi^2/s}\,.
\end{equation}
At higher energies, other intermediate states like $4 \pi$, $K \bar{K}$ etc.\ 
will contribute to the $s$-discontinuity of $A_{\mu \nu}$. Moreover, for a 
complete dispersive treatment one should also consider discontinuities in the 
$t$- and $u$-channels. We will not consider any of these contributions to the 
dispersion relation for $A_{\mu \nu}$, and we explain below on what grounds 
these approximations can be justified. 

%%%%%%%%%%%%%%%%%%%%%%%%%%%%%%%%%%%%%%%%%%%%%%%%%
\section{Dispersive framework for $K\to\pi\pi$}%%
\label{sec:Ktopipi}%%%%%%%%%%%%%%%%%%%%%%%%%%%%%%
%%%%%%%%%%%%%%%%%%%%%%%%%%%%%%%%%%%%%%%%%%%%%%%%%
The construction of a dispersion relation for $K_S\to\gamma\gamma^{(*)}$ 
requires input from $K_S\to\pi\pi$ and $\gamma\gamma^{(*)}\to\pi\pi$.
It is well known that one-loop chiral corrections to the $K_S\to\pi\pi$ 
amplitude are substantial, and largely due to significant rescattering 
effects of pions in the final state~\cite{Kambor:1991ah,Bertolini:1997ir,Pallante:2001he}. 
An understanding of FSI is thus essential in order to make sense of puzzles such 
as the $\Delta I =1/2$ rule or the SM prediction for $\epsilon'/\epsilon$.  As 
noted in Section \ref{sec:intro}, dispersive techniques are well suited to 
addressing FSI; here we review the dispersive framework 
\cite{Buchler:2001nm,Colangelo:2001uv} developed for $K\to\pi\pi$.

We begin with the standard isospin decomposition for the $K^0\to\pi\pi$ 
amplitude~\cite{Cirigliano:2011ny}
\begin{align}
  \frac{A_{\pi\pi}}{\sqrt{2}} &= A_{1/2}\,,
\end{align}
where $A_{1/2}$ is generated by the $\Delta I= 1/2$ component of ${\cal H}_w$, 
and we have omitted a term involving $\Delta I=3/2$.$^{\ref{rule}}$ As in 
Section \ref{sec:prelim}, we allow the effective weak Hamiltonian ${\cal H}_w$ 
to carry momentum $h_\mu \neq 0$, so the amplitude reads
\begin{equation}
  A_{1/2}(s,t',u') = 
  \langle (\pi(p_1)\pi(p_2))_{I=0} | {\cal H}_w^{1/2}(0)| K^0(k) \rangle\,,
\end{equation}
where the corresponding Mandelstam variables are given in (\ref{lhs cut}), and 
satisfy 
\begin{equation}
  s+t'+u' = 2\mpi^2 + \mk^2\,.
\end{equation}
The physical $K^0\to \pi\pi$ decay amplitude is then obtained by taking the 
limit $h_\mu \to 0$, at which point we have
\begin{equation}
  s=\mk^2 \quad \mbox{and} \quad t'=u'=\mpi^2\,.
  \label{Apipi ppt}
\end{equation}
If contributions from the imaginary parts of $D$ waves and higher are neglected, 
it is possible to decompose $A_{1/2}$ in terms of \emph{single-variable} 
functions
\begin{equation}
  A_{1/2}(s,t',u') = M_0(s) + C(s,t',u')\,,
  \label{Apipi dec}
\end{equation}
where the angular dependence is contained in
\begin{align}
  C(s,t',u') &= \frac{1}{3} \left[ N_0(t') + 2 R_0(t') \right]  \\
  &+ \frac{1}{2} \left[s-u' - \frac{\mpi^2(\mk^2-\mpi^2)}{t'} \right] N_1(t') 
  + \{t' \leftrightarrow u'\} \,, \nn
  \label{H fn}
\end{align}
and the explicit expressions for $N_i$ and $R_i$ can be found 
in~\cite{Buchler:2001nm}. 

As a result of this simplification, the dispersive treatment of the full
amplitude $A_{1/2}$ is reduced to solving a coupled set of dispersion
relations of the single-variable functions appearing in the right-hand side
of (\ref{Apipi dec}).  As shown in~\cite{Buchler:2001nm}, these relations
can be solved numerically, with a minimum of two subtraction
constants%
  \footnote{Constraints analogous to the Froissart-Martin 
  bound~\cite{Froissart:1961ux,Martin:1965jj} for two-particle scattering would 
  in principle allow even more subtractions. However, given the modest 
  information as regards the two we will be considering, this is currently a 
  purely academic question. The generous uncertainties assigned to the two 
  subtractions considered should also cover the possible presence of additional 
  subtraction constants.}
needed to ensure convergence of the dispersive integrals. One of these constants 
$a_{\pi\pi}$ can be determined at the soft-pion point 
\begin{equation}
  s=u'=\mpi^2 \quad \mbox{and} \quad t'=\mk^2\,,
  \label{spp}
\end{equation}
where $A_{1/2}$ is related to the on-shell $K\to\pi$ amplitude $A_\pi$:
\begin{align}
  -\frac{A_\pi}{2F_\pi} &= A_{1/2}(\mpi^2,\mk^2,\mpi^2) \nn \\
  &= a_{\pi\pi} +\frac{1}{3}\big[N_0(\mk^2)+2R_0(\mk^2)\big] + O(\mpi^2)\,. 
\end{align}
Note that with both $K$ and $\pi$ on-shell, the weak operator ${\cal H}_w$ in 
$A_\pi$ necessarily carries momentum.  The relevance of lattice calculations of 
$A_\pi$ in connection with the $\Delta I=1/2$ rule has recently been discussed 
in \cite{Crewther:2015dpa}.

On the other hand, the second constant $b_{\pi\pi}$ can be obtained by 
considering e.g.\ the derivative $\partial A_{1/2}/\partial s$ at the soft-pion 
point (\ref{spp}).  Ideally, lattice techniques would be used to determine 
$a_{\pi\pi}$ and $b_{\pi\pi}$, although such calculations remain to be 
undertaken.  Thus the approach taken in~\cite{Buchler:2001nm} was essentially 
pragmatic: to illustrate the role of FSI, the value of $b_{\pi\pi}$ was fixed by 
applying $\chi$PT$_3$, so that 
\begin{equation}
  b_{\pi\pi} = \frac{3a_{\pi\pi}(1+X)}{\mk^2-\mpi^2(4+3X)} + O(\mk^4)\,,
  \label{bpipi}
\end{equation}
where the dimensionless parameter $X$ controls the size of the expected NLO 
corrections; on the basis of the 3-flavour expansion it can be varied between 
$X=\pm 0.3$.  We note that the relation (\ref{bpipi}) is not affected by the 
weak mass term in ${\cal H}_w$; see Section \ref{sec:Kggstar ChPT}.

From the solutions to the dispersion relations, it is a straightforward matter 
to reconstruct the $K\to\pi\pi$ amplitude.  For $u'$ fixed near the physical 
value $\mpi^2$, it has been shown~\cite{Mercolli} that the contribution due to 
$C(s,t',u')$ is negligible relative to $M_0$ in the low-energy region 
$s \lesssim 1.5$ GeV$^2$.  Thus to a good approximation, we can write
\begin{align}
  A_{1/2}(s,\mk^2+\mpi^2-s,\mpi^2) \simeq 
  a_{\pi\pi}\big[ 1 + E(X) s/\mk^2 \big]\Omega_0^0(s)\,,
  \label{A12 disp}
\end{align}
where the quantity
\begin{equation}
  E(X) = \frac{3\mk^2(1+X)}{\mk^2-\mpi^2(4+3X)}
\end{equation}
parametrises the NLO corrections, and 
\begin{equation}
  \Omega_0^0(s) = \exp \Bigg( \frac{s}{\pi} \int_{4\mpi^2}^{\infty}\! 
  dz\, \frac{\delta_0^0(z)}{z(z-s-i\epsilon)} 
  \Bigg)
  \label{Omnes IJ}
\end{equation}
is the Omn\`{e}s function~\cite{Omnes:1958hv} subtracted at $s=0$, with 
$\delta_0^0$ the $\pi\pi$ scattering phase shift in the $I=\ell=0$ channel.

The $K\to\pi\pi$ amplitude in (\ref{A12 disp}) can be determined up to the 
unknown subtraction constant $a_{\pi\pi}$, modulo chiral corrections 
parametrised by $X$.  As a result, a first principles prediction for 
$K\to\pi\pi$ is not currently possible within this framework.  Fortunately, this 
does not pose a problem for $K_S\to\gamma\gamma^*$ since we can eliminate the 
dependence on $a_{\pi\pi}$ by matching to $A_{1/2}=A_0e^{i\delta_0}$ at the 
physical point (\ref{Apipi ppt}):
\begin{equation}
  |a_{\pi\pi}| = \frac{A_0}{|\Omega_0^0(\mk^2)| \big[ 1 + E(X) \big]}\,,
\end{equation}
where
\begin{equation}
  A_0 = (2.704\pm0.001)\times 10^{-7} \mbox{ GeV}
\end{equation}
is the empirical value of the $I=0$ amplitude~\cite{Cirigliano:2011ny}.  In this 
way, the dispersive representation of $K_S\to\gamma\gamma^*$ is largely 
determined in terms of measurable quantities, and as we show in 
Sections~\ref{sec:Ktogg}-\ref{sec:Ktogll} this leads to rather small 
uncertainties in our final results.

%%%%%%%%%%%%%%%%%%%%%%%%%%%%%%%%%%%%%%%%%%%%%%%%%%%%%%%%%%%
\section{Dispersion relations for $K_S \to \gamma\gamma$}%%
\label{sec:Ktogg}%%%%%%%%%%%%%%%%%%%%%%%%%%%%%%%%%%%%%%%%%%
%%%%%%%%%%%%%%%%%%%%%%%%%%%%%%%%%%%%%%%%%%%%%%%%%%%%%%%%%%%
As a first application of our dispersive framework, here we consider the case 
where both photons are on-shell.  A complete dispersive treatment of 
$K_S \to \gamma \gamma$ (with ${\cal H}_w$ carrying momentum) would require an 
analysis of all possible intermediate states in all three channels $s$, $t$ and 
$u$ --- clearly a daunting task. A simplification which has proven to be 
particularly effective for other scattering processes at low energies is 
to neglect the contributions to discontinuities coming from $D$ 
waves and higher. This leads to a dispersive representation of the scattering 
amplitude in terms of single-variable functions, much like in the case of the 
$K \to \pi \pi$ amplitude discussed in Section~\ref{sec:Ktopipi}. As in that 
case, we expect that at the physical point (\ref{phys pt}), the contributions to 
the $S$ wave coming from discontinuities in the $t$ and $u$ channels are 
negligible, and so will not consider them. Effectively this means that we 
construct a dispersion relation of the form-factor type (i.e.\ with a right-hand 
cut only), and only for the $S$ wave. Moreover, we will explicitly consider only 
the effect of $\pi \pi$ rescattering, which at low energies should be by far the 
most important one.  Indeed, this expectation is borne out by the LO 
$\chi$PT$_3$ result discussed in Section \ref{sec:Kggstar ChPT}.

Let us define $A_{\gamma\gamma}(s) = e^2 B_1(s)$, whose imaginary 
part coincides with the $s$-discontinuity in (\ref{ImA12}) once we set 
$q_2^2=0$:
\begin{align}
  &\imag_s\, A_{\gamma\gamma}(s) \nn \\
  &=-\alpha \frac{\sigma_\pi(s)}{\sqrt{2}s}
  \frac{\Omega_0^0(s)A_0}{|\Omega_0^0(\mk^2)|} 
  \frac{\big[ 1 + E(X)s/\mk^2 \big]}{\big[ 1 + E(X) \big]}[h_{0,++}^{0}(s)]^*\,. 
  \label{Im Agg}
\end{align}
Here $\alpha = e^2 /4\pi$ is the fine-structure constant, and $h_{0,++}^0$ 
is the projection of $h_{++}^0$ onto the $I=0$ channel. The real part then 
follows from a once-subtracted dispersion relation at $s=s_0$:
\begin{equation}
  A_{\gamma\gamma}(s) = a_{\gamma\gamma} + 
  \frac{s-s_0}{\pi}\int_{4\mpi^2}^\infty dz 
  \frac{\imag_s\, A_{\gamma\gamma}(z)}{(z-s_0)(z -s - i\epsilon)}\,, 
  \label{Re Agg}
\end{equation}
where $a_{\gamma\gamma}$ is the subtraction constant.  The subtraction is 
necessary because the $\pi^\pm,K^\pm$ loop contribution to the $\chi$PT$_3$ 
amplitude vanishes at the point
\begin{equation}
  s_0=-0.098 \mbox{ GeV}^{2}\,,
  \label{chi zero}
\end{equation}
and moreover to ensure convergence of the dispersive integral. This feature can 
be deduced from the explicit form of the $\chi$PT$_3$ amplitude in 
(\ref{Aggstar chpt}), with $q_2^2=0$. It follows that matching 
$A_{\gamma\gamma}(s_0)$ onto LO $\chi$PT$_3$ fixes $a_{\gamma\gamma}=0$, 
although in general, $a_{\gamma\gamma}$ will receive $SU(3)$ corrections due to 
terms at $O(p^6)$ in the chiral expansion. It is important to note that by 
matching below the $\pi\pi$ threshold, we make use of $\chi$PT$_3$ only in a 
kinematic region where the typically large corrections due to FSI are entirely 
absent, i.e.\ where the 3-flavour expansion should behave as expected.

To compute the integral in (\ref{Re Agg}), we require input for $h^0_{0,++}$ and 
the $S$ wave of the $K_S \to \pi \pi$ amplitude which, in our representation, is 
given by $\Omega_0^0$. Concerning the latter, the dispersive representation of 
the single-variable functions in (\ref{Apipi dec}) is only valid in the elastic 
scattering region $4 \mpi^2 < s < 16\mpi^2$, even though the first significant 
inelastic contribution is due to the $K\bar{K}$ intermediate state when 
$s > 4 m_K^2$. Taking this into account would require a coupled-channel analysis 
of $K_S\to\pi\pi$ and $K_S\to KK$, which is beyond the scope of this work. 
Moreover, it is unclear whether this would lead to better precision, because 
there are no sources of experimental information on $K_S \to KK$, and we would 
have to rely completely on $\chi$PT$_3$ to determine the subtraction constants,
with correspondingly large uncertainties.

\begin{figure}[t]
  \centering
  \subfloat{\includegraphics[scale=0.65]{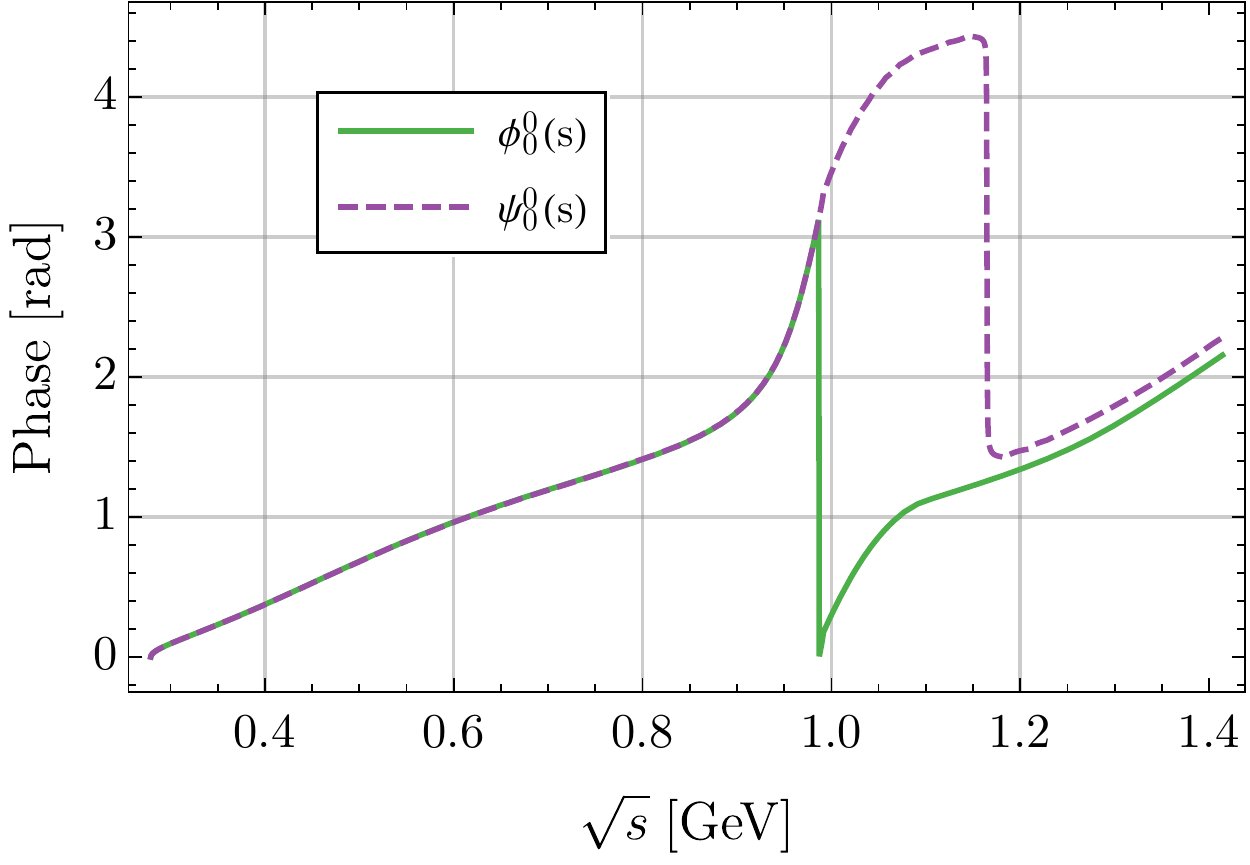}}
  \quad
  \subfloat{\includegraphics[scale=0.65]{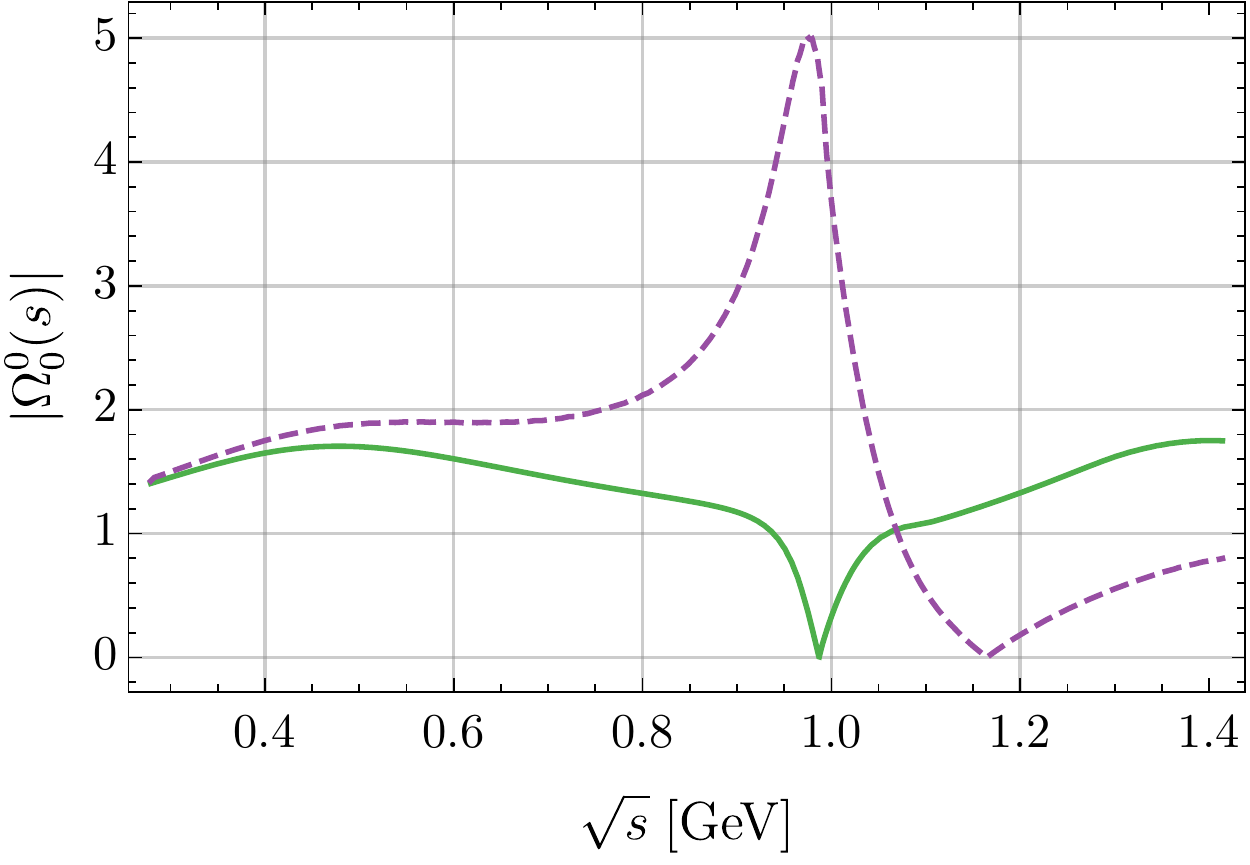}}
  \caption{Energy dependence of phase shift inputs (top) and magnitude of 
  the corresponding Omn\`{e}s functions (bottom).} \label{fig:phase_omnes}
\end{figure}

We will thus stick to a single-channel treatment and only consider the 
contribution to the imaginary part of the $S$ wave specified in 
(\ref{Im Agg}). This implies that the phases of $h^0_{0,++}$ and $\Omega_0^0$ 
have to match exactly in order for $\imag_s\, A_{\gamma\gamma}$ to be 
real, as it should be in a single-channel treatment. This is guaranteed in the 
elastic region, which effectively extends up to $s=4 m_K^2$, but above that 
threshold an ambiguity arises: do the phases of the $K_S\to\pi\pi$ partial
waves continue to behave like the elastic scattering phase shifts
$\delta_\ell^I$, or do they exhibit a sharp ``dip'' like the one 
observed~\cite{Ananthanarayan:2004xy} in the scalar form factor of the pion?%
  \footnote{See also the discussion in \cite{Hoferichter:2012wf} which shows 
  how, in the coupled-channel treatment, the phase of the scalar form factor of 
  the pion is sensitive to the input for the subtraction constants. We stress, 
  however, that even in cases where the phase of the form factor continues to 
  track $\delta_0^0$ after the $K \bar{K}$ threshold, the modulus of the 
  form factor still has a dip rather than a peak at $s=4 M_K^2$.} 
This ambiguity affects both quantities: $h^0_{0,++}$ as well as $\Omega_0^0$. We 
take a pragmatic approach to the problem and follow 
Moussallam~\cite{Moussallam:2013una}, who constructs a phase with the property
\begin{equation}
  \phi_0^0(s) = \left\{ \begin{array}{lcc} 
  \delta_0^0(s)\,, & \qquad& s \leq s_\pi \\
  \delta_0^0(s) - \pi\,, & \qquad & s > s_\pi
  \end{array} \right.\,,
 \label{eq:phi00}
\end{equation}
where $s_\pi$ lies near the $K\bar K$ threshold and is the point where 
$\delta_0^0$ crosses $\pi$. The corresponding Omn\`{e}s function 
$\Omega_0^0[\phi]$ thus displays a ``dip'' across the inelastic region. Another 
option is to evaluate the Omn\`es function with the phase of $h^0_{0,++}$, 
\begin{equation}
  \psi_0^0(s) = \arg h_{0,++}^0(s)\,,
\label{eq:psi00}
\end{equation}
as input. Watson's theorem ensures $\phi_0^0 = \psi_0^0$ in the elastic
region, and leads to two representations for $\Omega_0^0$ which are in very 
close agreement at low energy. A comparison of the two phases and
corresponding Omn\`{e}s factors is shown in Figure \ref{fig:phase_omnes}.

As argued by 
Moussallam~\cite{Moussallam:2013una} and earlier by Morgan and 
Pennington~\cite{Morgan:1983re} (see also the discussion in 
\cite{Ananthanarayan:2004xy}), unless the operator which is responsible for the
creation of the pion pair has a large overlap with the $f_0(980)$, one expects a 
weak coupling to the $f_0(980)$, and correspondingly a dip in the amplitude. The 
only known example of an operator whose amplitude would have a peak instead of a 
dip is that of the $\bar{s} s$ operator. 

We thus conclude that our preferred phase is the one given in Eq.~(\ref{eq:phi00}) 
and with this we will obtain our central results. The phase in (\ref{eq:psi00}) 
will be used to estimate our systematic uncertainty. More extreme behaviours -- 
like ``Solution 1'' in \cite{Colangelo:2015kha} -- are deemed to be very
unlikely and will not be considered.

\begin{figure}[t]
  \centering
  \includegraphics[scale=0.65]{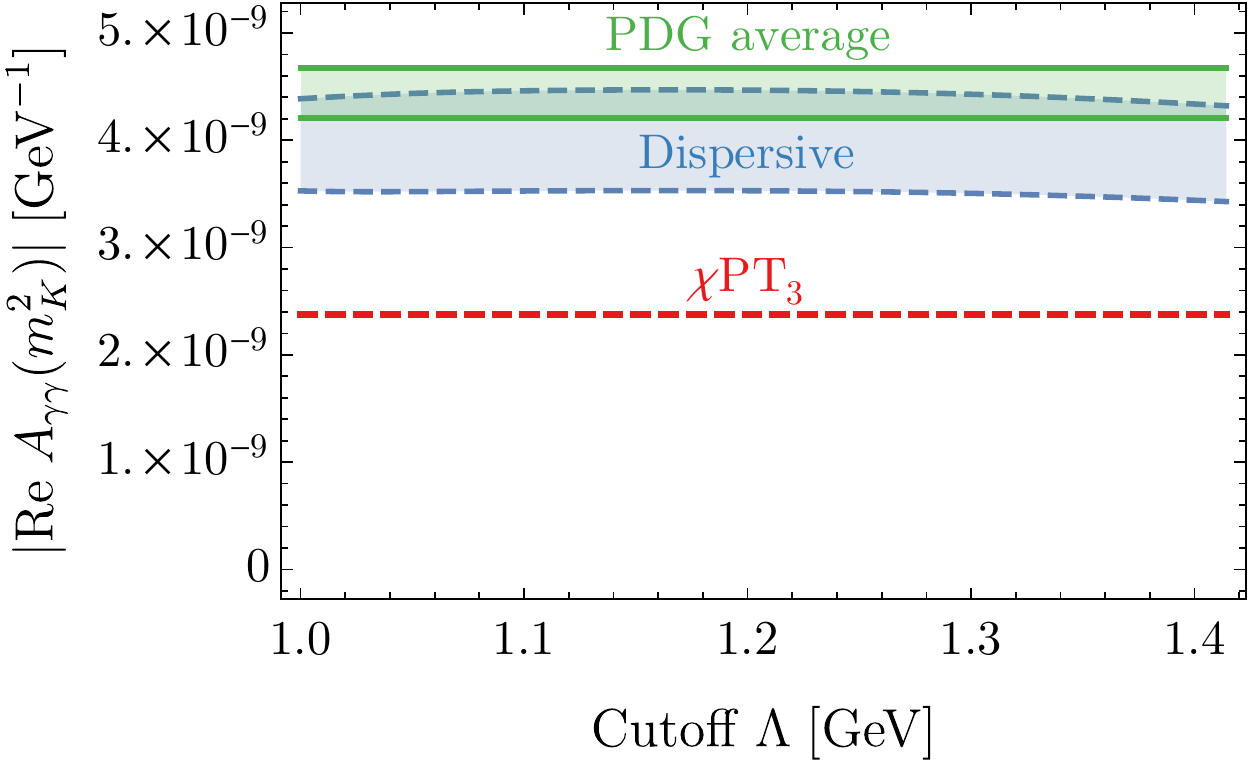}
  \caption{Cutoff dependence of the dispersive amplitude 
  $|\mbox{Re }A_{\gamma\gamma}|$ at the physical point (\ref{phys pt}), where 
  the blue band corresponds to the systematic uncertainty. For comparison, the 
  PDG value of $|\mbox{Re } A_{\gamma\gamma}|$ and its $1\sigma$ uncertainties 
  is shown by the green band, while the lowest order prediction from 
  $\chi$PT$_3$ is shown by the dashed red line.} 
  \label{fig:abs_Agg}
\end{figure}

Regarding the input for $h_{0,++}^0$, we use data from the coupled-channel 
analysis of $\gamma\gamma\to\pi\pi$ performed by Garc\'{i}a-Mart\'{i}n and 
Moussallam (GMM)~\cite{GarciaMartin:2010cw}. Since the determination of 
$h_{0,++}^0$ in this analysis is expected to be reliable up to $s \lesssim 2$ 
GeV$^2$,%
  \footnote{B.~Moussallam, private communication.}
it is necessary to impose a cutoff $\Lambda$ in our dispersion integral 
(\ref{Re Agg}). At the physical point $s=\mk^2$, a comparison of the cutoff 
dependence is shown in Figure \ref{fig:abs_Agg}, where 
$|\mbox{Re }A_{\gamma\gamma}|$ is seen to exhibit a very mild sensitivity to 
variations in $\Lambda$.

\begin{figure}[t]
  \centering
  \hspace{-3mm}\subfloat{\includegraphics[scale=0.65]{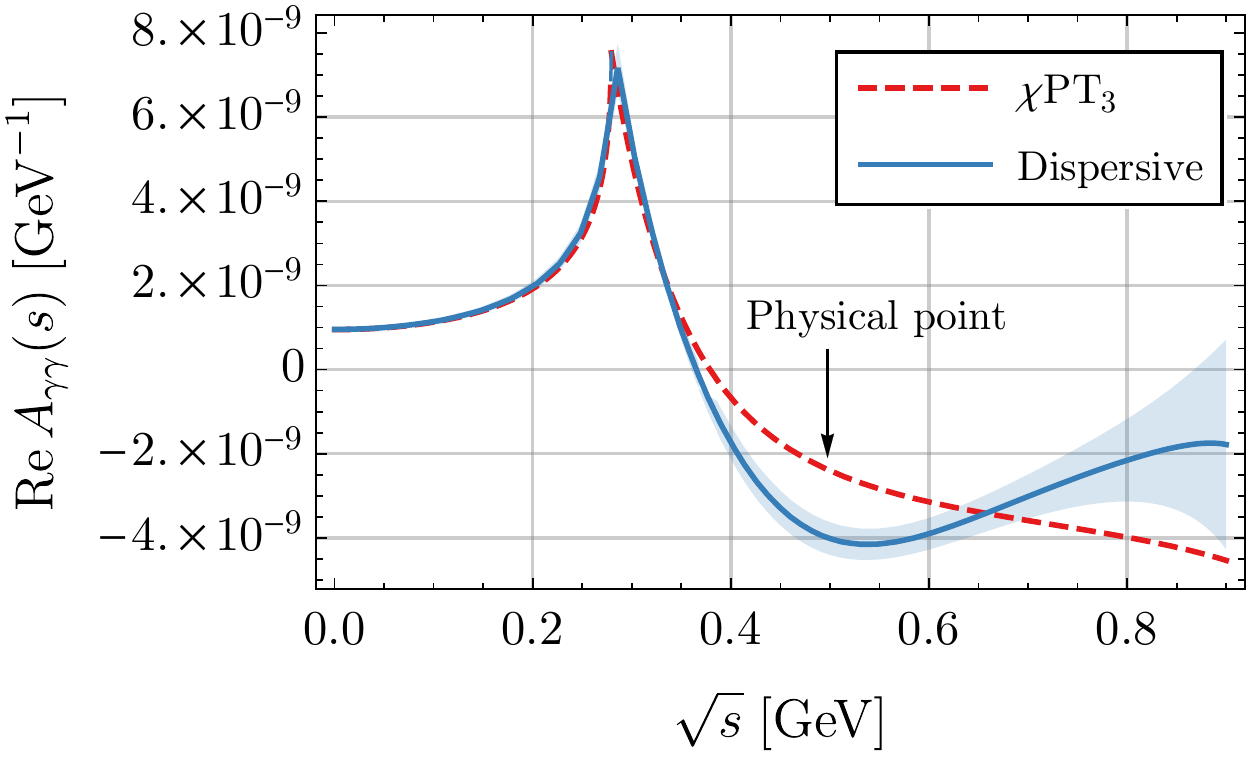}}
  \quad
  \subfloat{\includegraphics[scale=0.65]{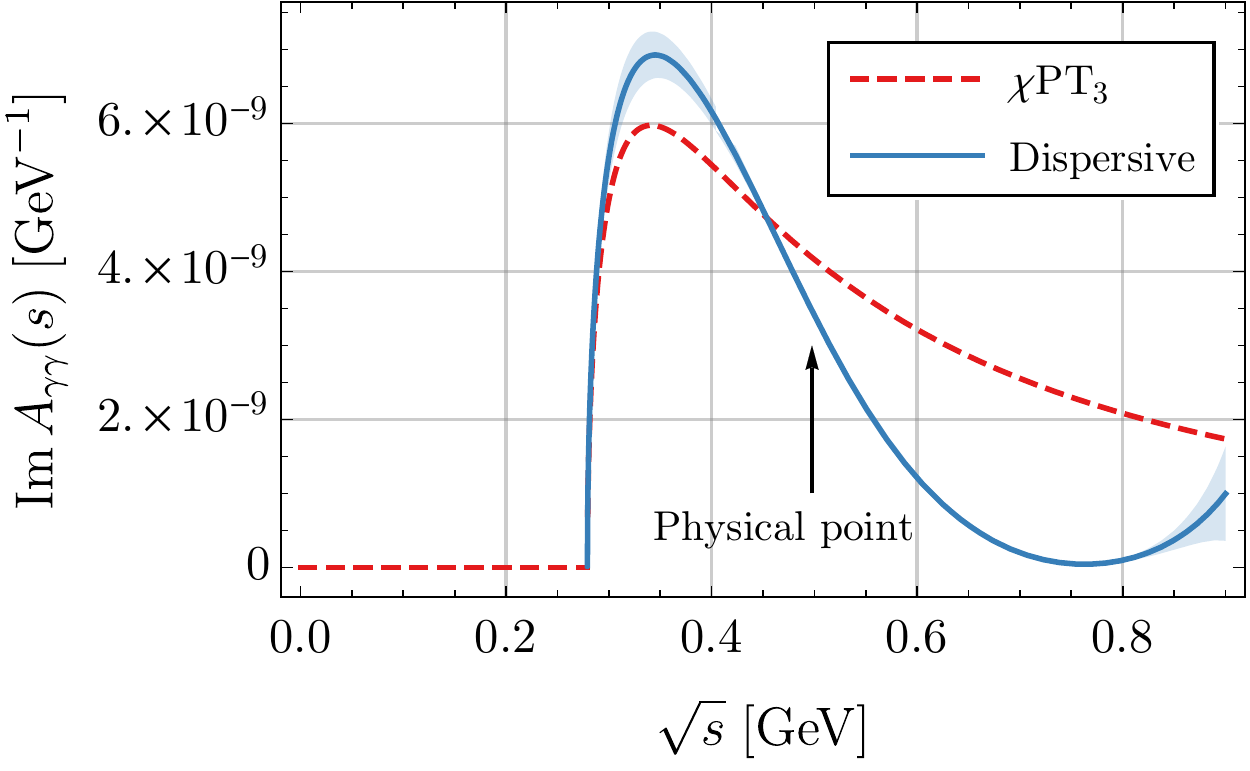}}
  \caption{Energy dependence of the real and imaginary parts of the 
  $K_S\to\gamma\gamma$ amplitude.  The blue band in the dispersive result 
  corresponds to the systematic uncertainty.}
  \label{fig:Re_Im_Agg}
\end{figure}

Taking $\Lambda = 1.2$ as a benchmark value, the energy dependence of the real 
and imaginary parts of $A_{\gamma\gamma}$ is shown in 
Figure~\ref{fig:Re_Im_Agg}.  As expected, the dispersive representation agrees 
with LO $\chi$PT$_3$ below the $\pi\pi$ threshold.  However, for $s > 4\mpi^2$, 
the effects from FSI distort the amplitude, producing a significant enhancement 
(suppression) of the real (imaginary) part. These effects lead to an enhanced 
prediction for the branching ratio
\begin{align}
  \mathrm{BR}(K_S \to \gamma\gamma) 
  &= \frac{\mk^3}{64\pi} 
  \frac{|A_{\gamma\gamma}(\mk^2)|^2}{\Gamma(K_S)_\mathrm{tot}} \nn \\
  &=(2.34 \pm 0.31) \times 10^{-6}
\end{align}
which brings the SM and experiment (\ref{BR Kgg}) into much better agreement.
The uncertainty has been determined by considering the variation $X=\pm 0.3$, 
shifting the value of $s_0$ by 30\%, the comparison of the two Omn\`{e}s inputs 
(Figure \ref{fig:phase_omnes}), and an estimate of contributions from the 
high-energy region $\Lambda > 1.2$ GeV, where the phase of $\Omega_0^0$ is 
guided to $\pi$ and the helicity partial wave is fixed to a constant value 
$|h_{0,++}| \approx 4$. Combined in quadrature, the final uncertainty has turned 
out to be remarkably modest.

%%%%%%%%%%%%%%%%%%%%%%%%%%%%%%%%%%%%%%%%%%%
\subsection{Comparison to the literature}%%
%%%%%%%%%%%%%%%%%%%%%%%%%%%%%%%%%%%%%%%%%%%
As shown in Figure \ref{fig:Re_Im_Agg}, the real part of $A_{\gamma\gamma}$ 
receives a significant enhancement in absolute value at $s=\mk^2$ due to FSI.  A 
similar observation has been made by Kambor and Holstein (KH) 
\cite{Kambor:1993tv}, who estimated the effects of $\pi\pi$ rescattering in 
$K_S\to\gamma\gamma$ and $K_L\to\pi^0\gamma\gamma$ by extrapolating the kaon 
mass off-shell.  Focusing on the former process, we can adapt their notation to 
ours by defining 
\begin{equation}
  A_{\gamma\gamma}^\mathrm{KH}(s) = - 2\alpha F_\pi B(s) / s\,,
  \label{rescale}
\end{equation}
where $B(s)$ is a scalar function whose definition is given 
in~\cite{Kambor:1993tv}. In our comparison, we have updated the input used 
in~\cite{Kambor:1993tv} to account for improved 
determinations~\cite{GarciaMartin:2010cw} of the Omn\`{e}s factor and helicity 
partial wave. The resulting predictions at the benchmark value of $\Lambda=1.2$ 
GeV are shown in Table~\ref{table:Agg pred}, where we also list the pure 
octet$^{\ref{rule}}$ predictions from $\chi$PT$_3$.  We note that  although the 
KH formalism produces a branching ratio consistent with experiment, it relies on 
the assumption that one can extrapolate the kaon mass off the mass shell.  As 
discussed in~\cite{Colangelo:2001uv,Buchler:2001np}, this procedure suffers from 
an inherent ambiguity as there is \emph{no unique way} in which to perform the 
off-shell extrapolation.  By contrast, our framework always involves on-shell 
states, and is free from such ambiguities.  
\begin{table*}[hbt]
\centering 
\begin{tabular}{l c c c c} 
\hline\hline\bigstrut
Input & Re$\,A_{\gamma\gamma}$ $\left[10^{-9} \mbox{ GeV}^{-1}\right]$ 
& $\Im\,A_{\gamma\gamma}$ $\left[10^{-9} \mbox{ GeV}^{-1}\right]$ 
& $|A_{\gamma\gamma}| \left[10^{-9} \mbox{ GeV}^{-1}\right]$ 
& BR$(K_S\to\gamma\gamma)$ $[10^{-6}]$ \bigstrut \\ 
\hline\bigstrut 
$\chi$PT$_3$ & $-2.38$ & 4.19 & 4.82 & $1.9$  \\  
KH & $-4.28$ & 3.47 & $5.51$ & $2.54$ \\
This work & $-4.00\pm 0.47$ & $3.47$ & $5.30\pm 0.35$ & $2.34 \pm 0.31$ \\ 
PDG & -- & -- & $5.62 \pm 0.18$ & $2.63\pm 0.17$  \\
\hline \hline 
\end{tabular} 
\caption{
Determinations of the $K_S\to\gamma\gamma$ amplitude $A_{\gamma\gamma}$ 
and branching ratio at the physical point $s=\mk^2$. The numbers in the 
row labelled ``This work'' have been obtained with input from GMM.}
\label{table:Agg pred}
\end{table*}

%%%%%%%%%%%%%%%%%%%%%%%%%%%%%%%%%%%%%%%%%%%%%%%%%%%%%%%%%%%%%%%%%
\section{Dispersion relations for $K_S \to \gamma\ell^+\ell^-$}%%
\label{sec:Ktogll}%%%%%%%%%%%%%%%%%%%%%%%%%%%%%%%%%%%%%%%%%%%%%%%
%%%%%%%%%%%%%%%%%%%%%%%%%%%%%%%%%%%%%%%%%%%%%%%%%%%%%%%%%%%%%%%%%
We now consider the case where the photon momentum in $K_S\to\gamma\gamma^*$ 
can remain off-shell $q_2^2\neq 0$. As in Section 
\ref{sec:Ktogg}, we focus on contributions from $S$ waves and define 
$A_{\gamma\gamma^*}(s,q_2^2) = e^2 B_1(s,q_2^2)$.

In the presence of $\pi\pi$ rescattering in the $I=0$ channel, the 
$s$-discontinuity reads
\begin{align}
  &\mathrm{disc}_s \, A_{\gamma\gamma^*}(s,q_2^2) \\
  &=-\alpha \frac{\sigma_\pi(s)}{\sqrt{2}}
  \frac{\Omega_0^0(s)A_0}{|\Omega_0^0(\mk^2)|} 
  \frac{\big[ 1 + E(X)s/\mk^2 \big]}{\big[ 1 + E(X) \big]} 
  \frac{[h_{0,++}^{0}(s,q_2^2)]^*}{s-q_2^2}\,, \nn
\end{align}
so the corresponding dispersion integral is given by%
  \footnote{The absence of anomalous thresholds in $K_S\to\gamma\gamma^*$ 
  follows from the same arguments used for 
  $\gamma\gamma^*\to\pi\pi$~\cite{Moussallam:2013una}; see 
  also~\cite{Hoferichter:2013ama} for a general treatment.}
\begin{equation}
  A_{\gamma\gamma^*}(s,q_2^2) = a_{\gamma\gamma^*}(q_2^2) 
  + \frac{s}{\pi} \int_{4\mpi^2}^{\infty} dz \,
  \frac{\mathrm{disc}_s \,A_{\gamma\gamma^*}(z,q_2^2)}{z(z-s-i\epsilon)}\,,
  \label{Aggstar sub zero}
\end{equation} 
where we have subtracted at $s_0=0$ to ensure convergence of the dispersive 
integral, and fixed the subtraction constant by 
matching to the $\chi$PT$_3$ amplitude (\ref{Aggstar chpt}): 
\begin{equation}
  a_{\gamma\gamma^*}(q_2^2) = e^2B_1(0,q_2^2)\big|_{{\chi\mathrm{PT}_3}} 
  \equiv A_{\gamma\gamma^*}(0,q_2^2)\big|_{{\chi\mathrm{PT}_3}}\,.
\end{equation}

To evaluate (\ref{Aggstar sub zero}), we begin by decomposing the helicity 
partial wave 
\begin{equation}
  h_{0,++}^0(s,q_2^2) = h_{0,++}^{0,\mathrm{Born}}(s,q_2^2) 
  + h_{0,++}^{0,\mathrm{scatt}}(s,q_2^2)\,,
\end{equation}
noting that Low's theorem~\cite{Low:1958sn} implies the Born-subtracted partial 
wave $h_{0,++}^{0,\mathrm{scatt}}$ has a zero at $s=q_2^2$ (i.e.\ when the 
on-shell photon becomes soft $q_1 \to 0$). 

The Born contribution to the helicity partial wave%  
  \footnote{The Clebsch-Gordan factor of $\sqrt{4/3}$ is due to the 
  rotation from the charge basis to the isospin one \cite{Moussallam:2013una}.}
\begin{equation}
  h_{0,++}^{0,\mathrm{Born}}(s,q_2^2)= 
  -\sqrt{\frac{4}{3}}\frac{F_\pi^V(q_2^2)}{s-q_2^2} 
  \bigg[ \frac{4 \mpi^2}{\sigma_\pi(s)} 
  \ln \dfrac{1+\sigma_\pi(s)}{1-\sigma_\pi(s)} - 2q_2^2 \bigg]
\end{equation}
produces a double pole $\sim (s-q_2^2)^2$ in 
$\mathrm{disc}_s\, A_{\gamma\gamma^*}$, so a decomposition of the 
integrand
\begin{align}
  &\frac{1}{(z-s)(z-q_2^2)^2} \nn \\
  &= 
  \frac{1}{(s-q_2^2)^2} \left[ \frac{1}{z-s} - \frac{1}{z-q_2^2}\right] 
  - \frac{1}{s-q_2^2} \frac{1}{(z-q_2^2)^2} \,, 
  \label{id1}
\end{align}
is required in order to evaluate the dispersive integral numerically.  In the 
above, $F_\pi^V$ denotes the vector form factor of the pion, and is set to unity 
in LO $\chi$PT$_3$. Using the identity in (\ref{id1}), we get the Born part of 
the $K_S\to\gamma\gamma^*$ amplitude
\begin{align}
  &A_{\gamma\gamma^*}^\mathrm{Born}(s,q_2^2) \label{Aggstar born}\\
  &= \frac{s}{\pi}\left\{ \frac{Q(s,q_2^2) - Q(q_2^2,q_2^2)}{(s-q_2^2)^2} 
  - \frac{1}{s-q_2^2} \left[ 
  \frac{\partial}{\partial\lambda} Q(\lambda,q_2^2) \right]_{\lambda=q_2^2} 
  \right\}\,, \nn
\end{align}
where we have defined
\begin{align}
  Q(s,q^2) &= \int_{4\mpi^2}^{\infty} dz 
  \frac{(z-q^2)^2\, \mathrm{disc}_s \, 
  A_{\gamma\gamma^*}^\mathrm{Born}(z,q^2)}{z(z-s-i\epsilon)}\,.
\end{align}
Similarly, for the rescattering contribution, we use the identity
\begin{equation}
  \frac{1}{(z-s)(z-q_2^2)} = \frac{1}{s-q_2^2} 
  \left[ \frac{1}{z-s} - \frac{1}{z-q_2^2} \right]
\end{equation}
so that
\begin{equation}
  A^{\mathrm{scatt}}_{\gamma\gamma^*}(s,q_2^2) = \frac{s}{\pi} 
  \left\{\frac{R(s,q_2^2)-R(q_2^2,q_2^2)}{s-q_2^2}\right\}\,, 
  \label{Aggstar scatt}
\end{equation}
where
\begin{align}
  R(s,q^2) &= \int_{4\mpi^2}^{\infty} dz 
  \frac{(z-q^2)\,\mathrm{disc}_s \, 
  A_{\gamma\gamma^*}^\mathrm{scatt}(z,q^2)}{z(z-s-i\epsilon)}\,.
\end{align}

\begin{figure}[b]
  \centering
  \subfloat{\includegraphics[scale=0.6]{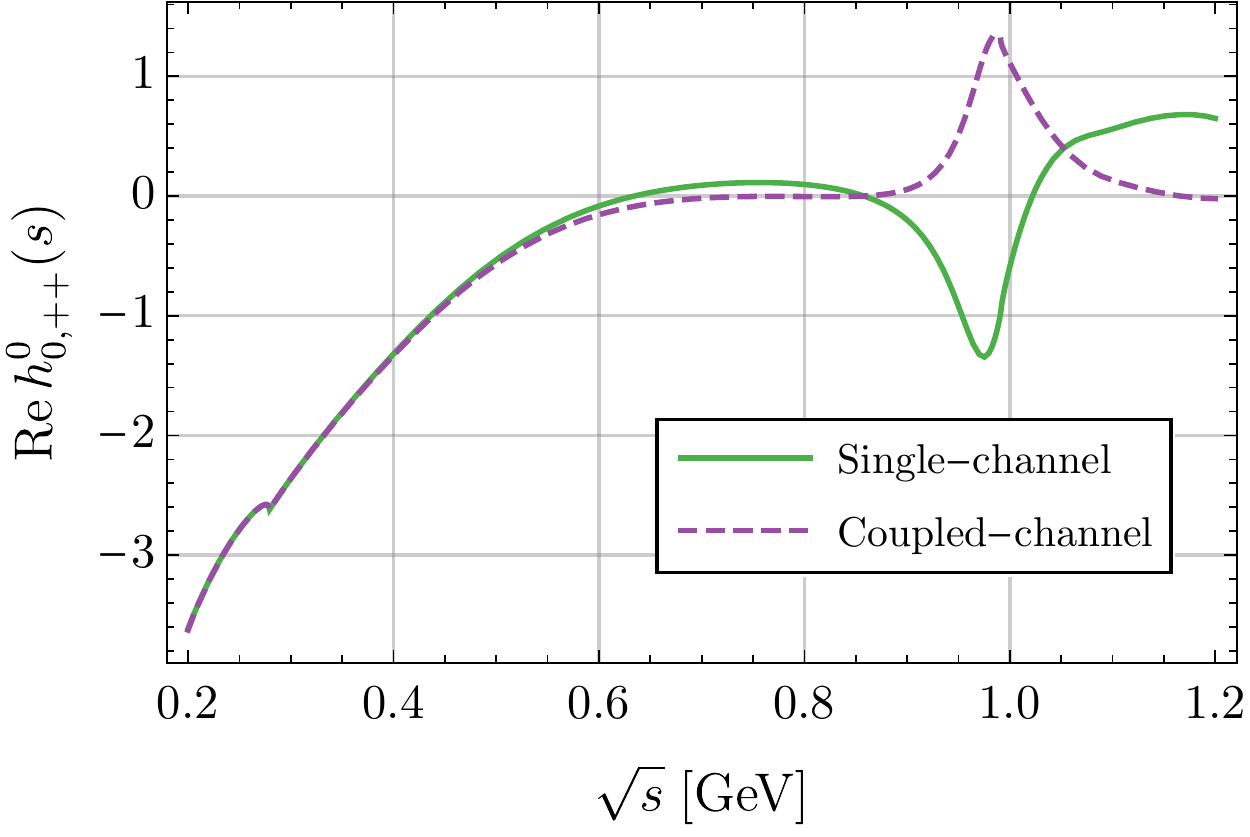}} \;
  \subfloat{\includegraphics[scale=0.6225]{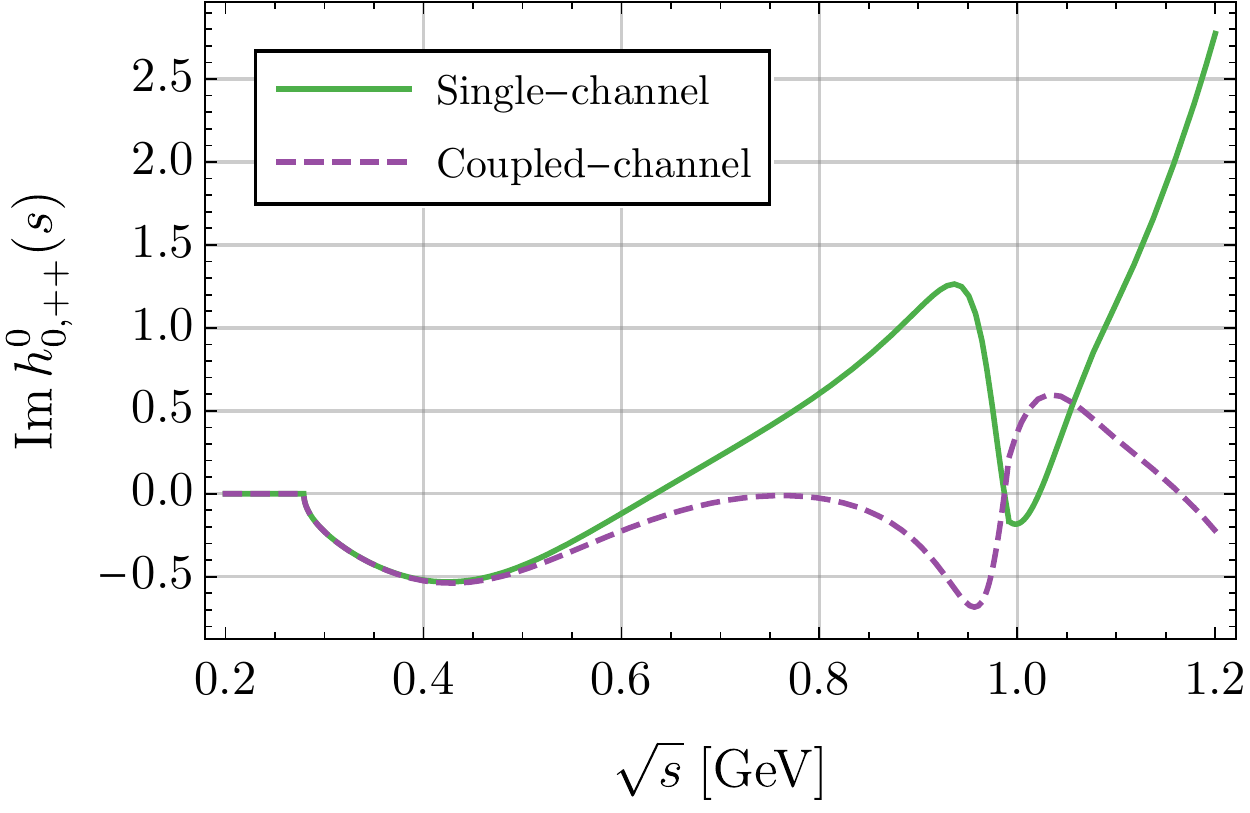}}
  \caption{Energy dependence of helicity partial waves obtained from dispersive 
  analyses of $\gamma\gamma^{(*)}\to\pi\pi$~\cite{GarciaMartin:2010cw,Moussallam:2013una}.}  
  \label{fig:h00 single coupled}
\end{figure}

\begin{figure*}[ht]
  \centering
  \subfloat{\includegraphics[scale=0.45]{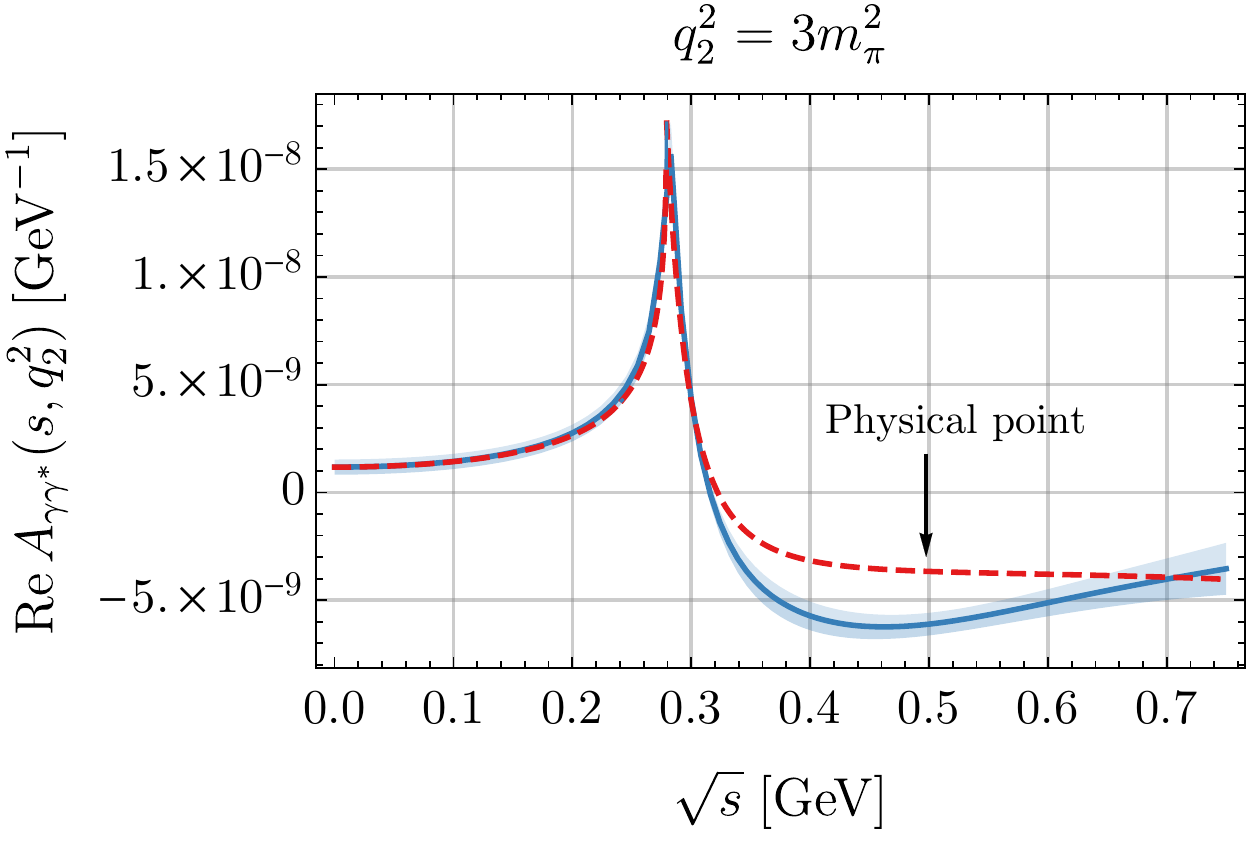}}
  \quad
  \subfloat{\includegraphics[scale=0.45]{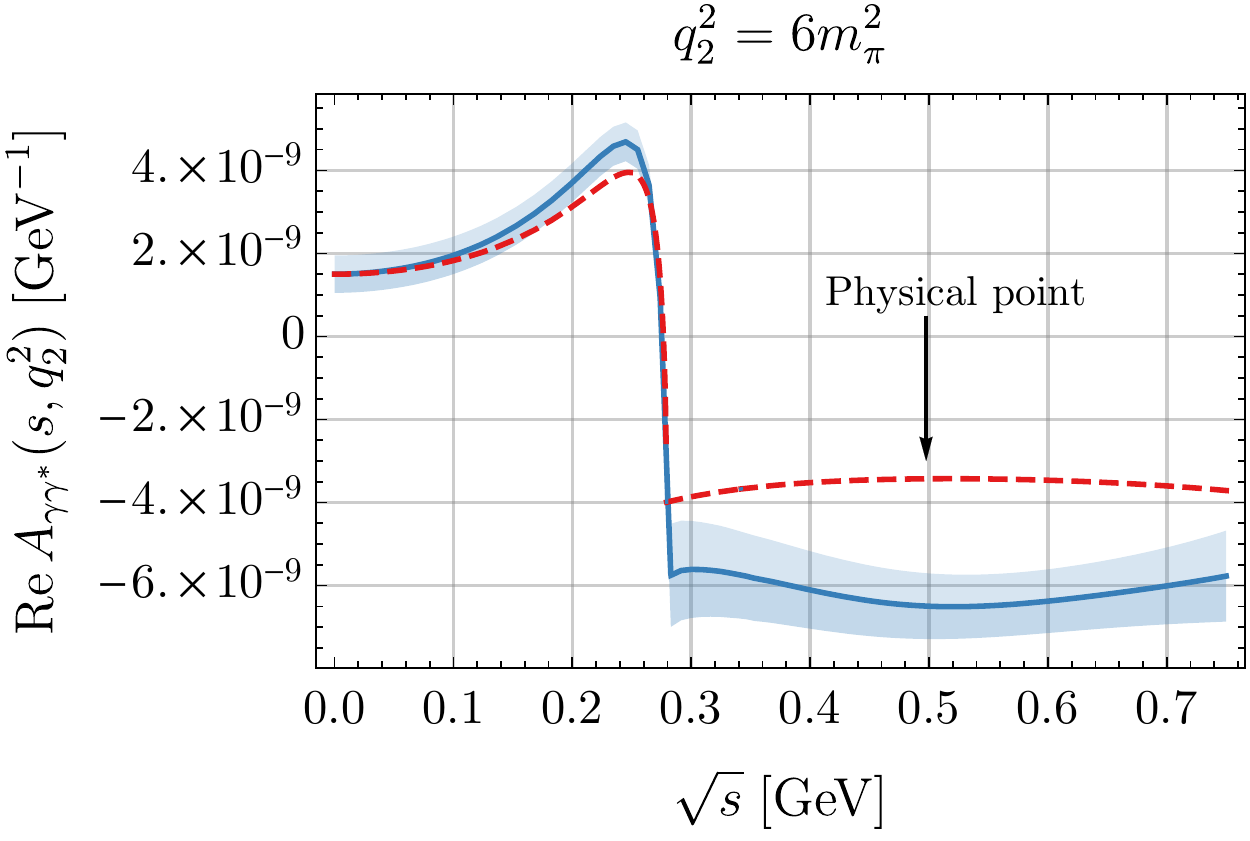}} 
  \quad
  \subfloat{\includegraphics[scale=0.45]{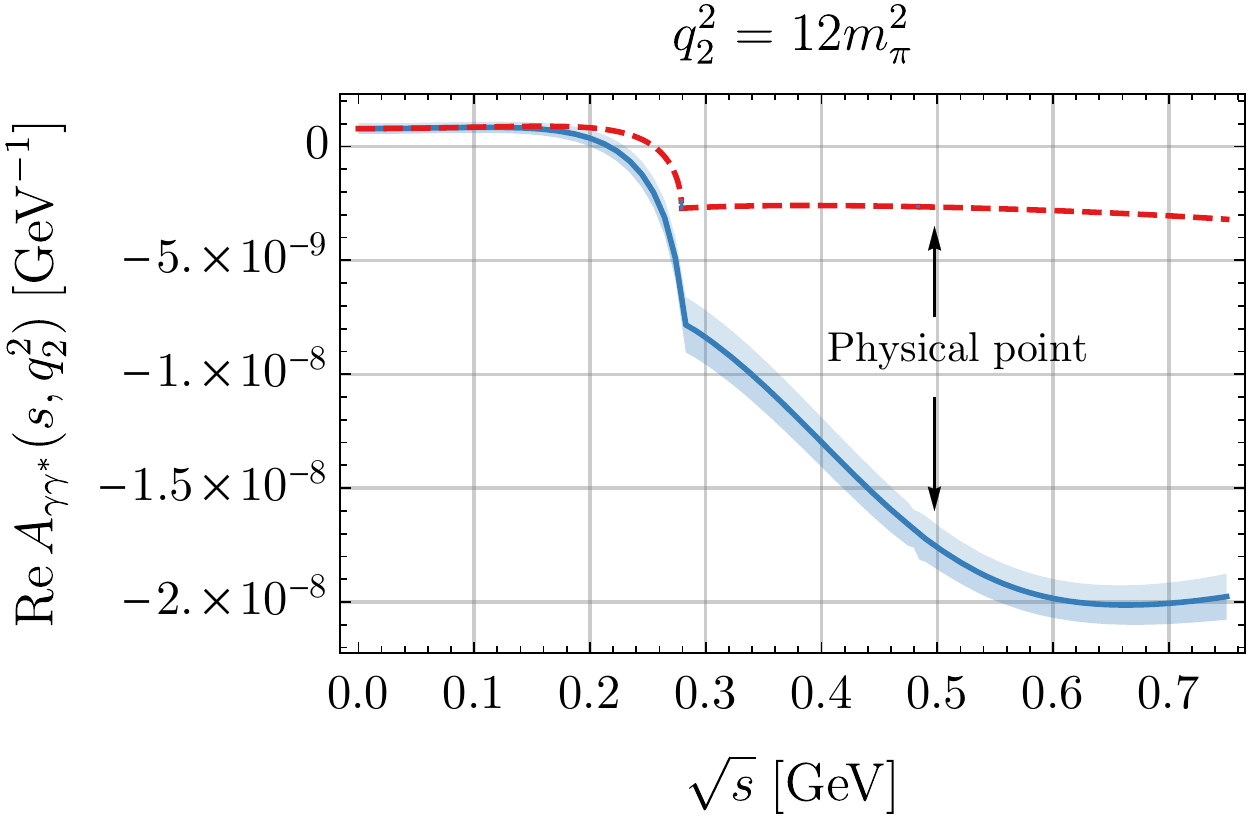}}
  \quad
  \subfloat{\includegraphics[scale=0.45]{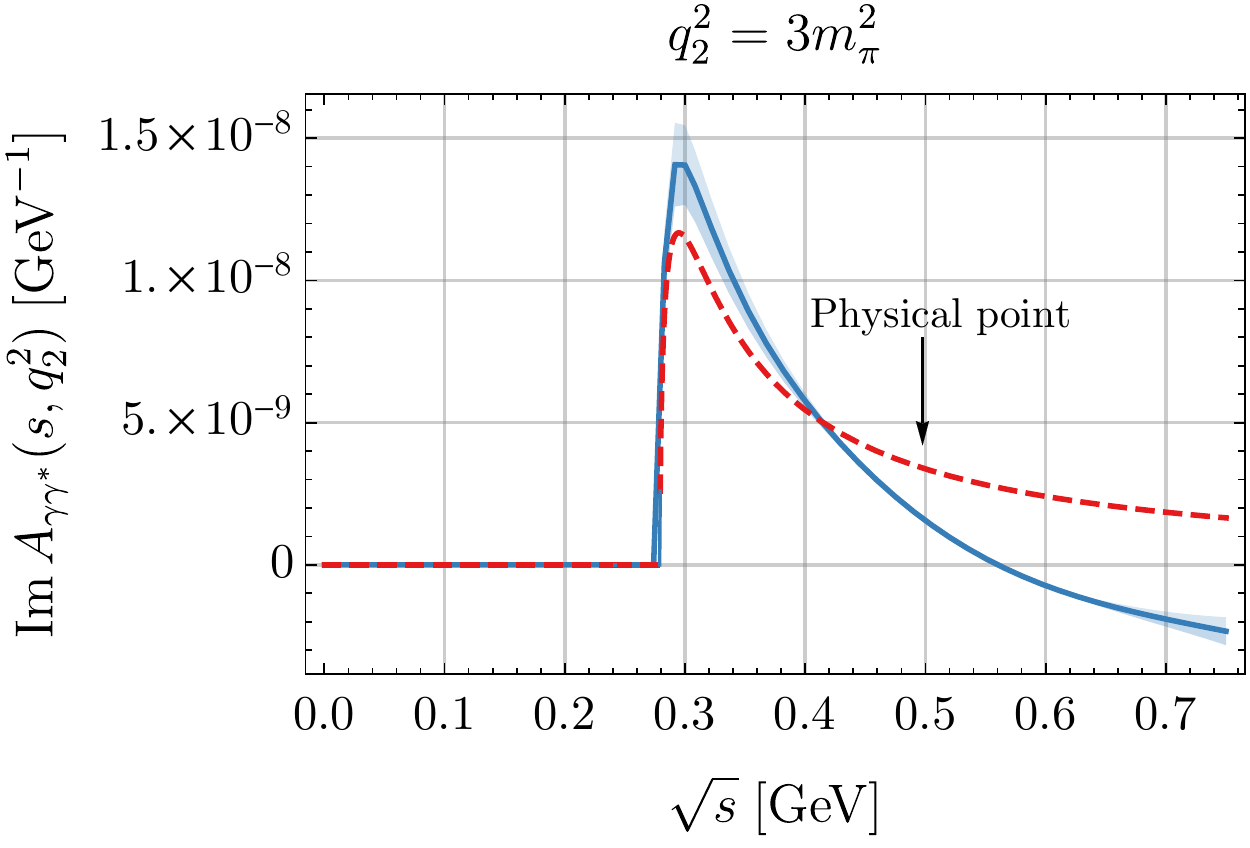}}
  \quad
  \subfloat{\includegraphics[scale=0.45]{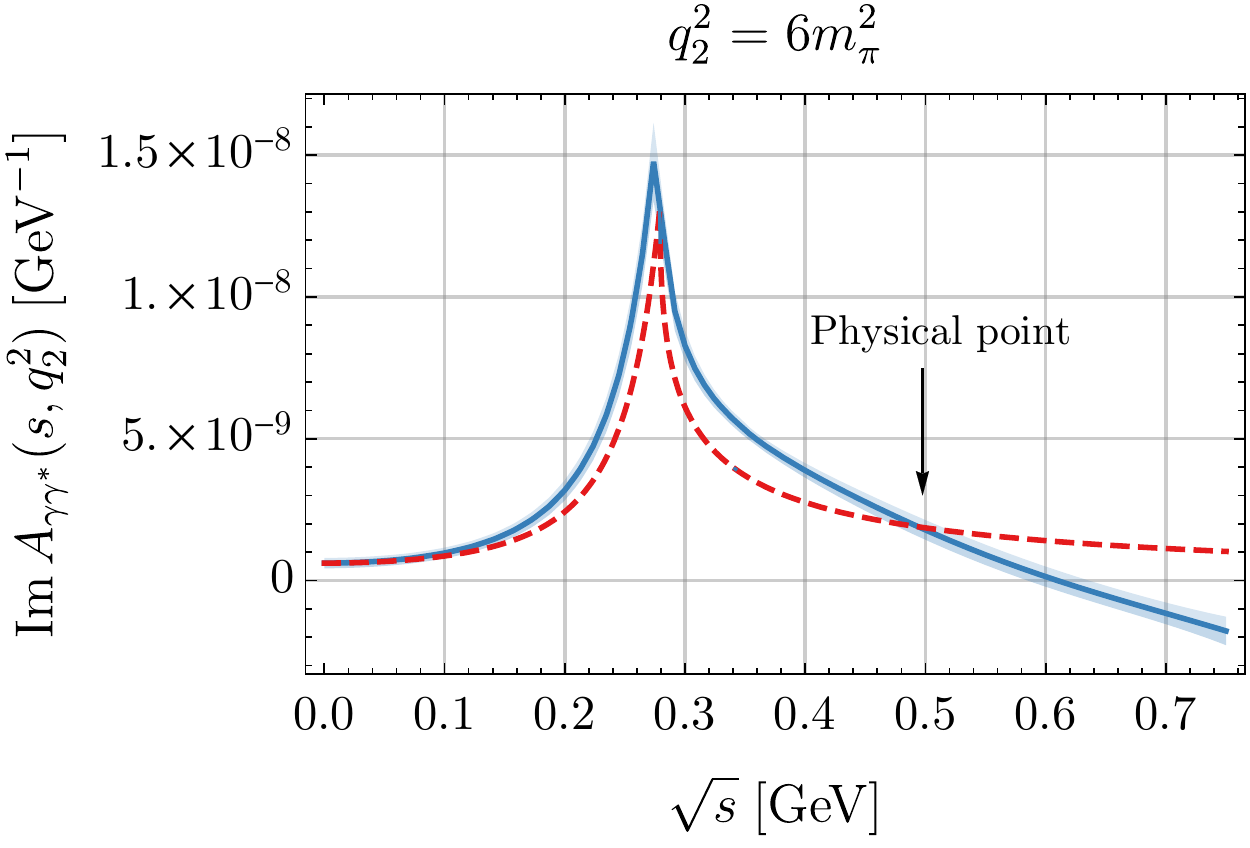}} 
  \quad
  \subfloat{\includegraphics[scale=0.45]{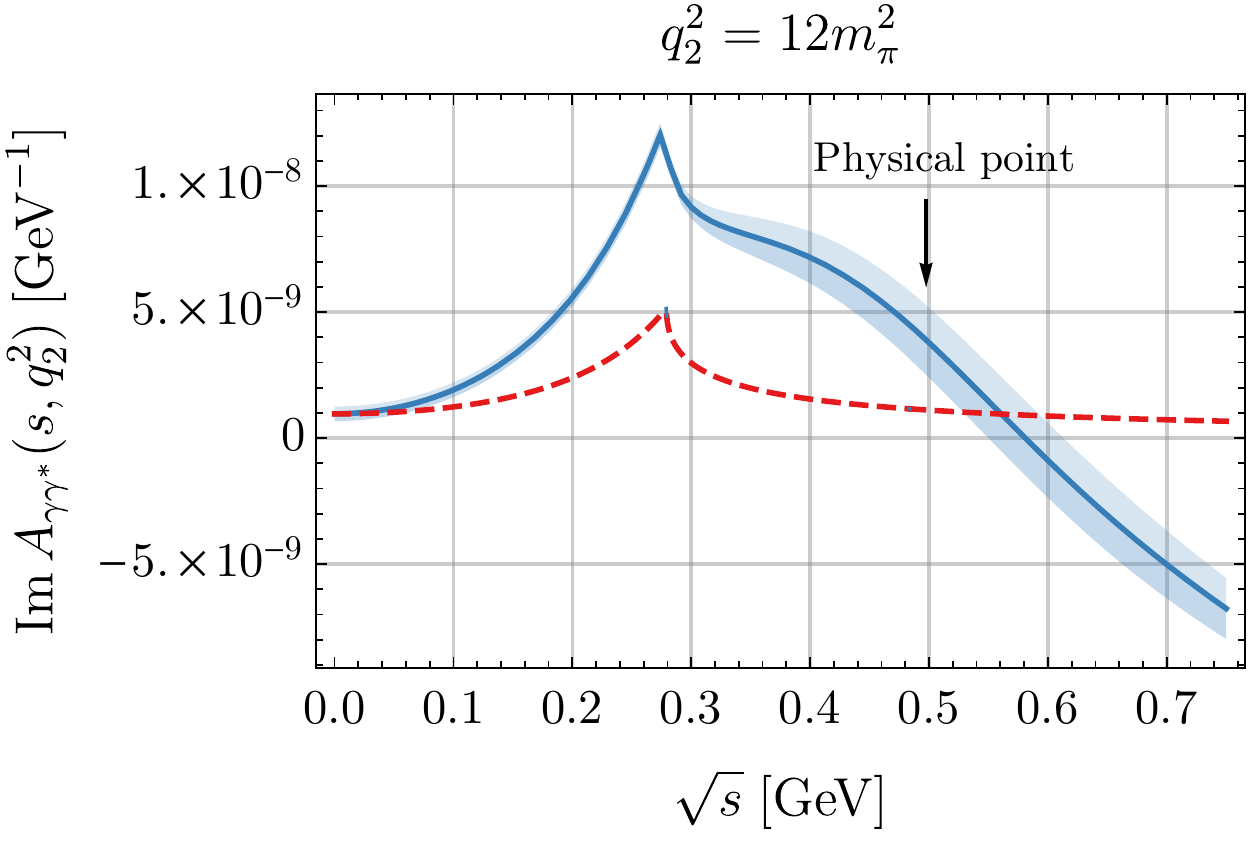}}
  \caption{Energy dependence of the real and imaginary parts of the 
  $K_S\to\gamma\gamma^*$ amplitude $A_{\gamma\gamma^*}$ for fixed values of 
  $q_2^2$. Colour coding as in Figure \ref{fig:Re_Im_Agg}.} 
  \label{fig:Aggstar}
\end{figure*}

In the evaluation of (\ref{Aggstar born}) and (\ref{Aggstar scatt}), we use the 
two Omn\`{e}s inputs discussed in Section \ref{sec:Ktogg}, as well as the 
pion form factor and helicity partial waves $h_{0,++}^{0}$ obtained from 
Moussallam's single-channel analysis of 
$\gamma\gamma^*\to\pi\pi$~\cite{Moussallam:2013una}.  The range of validity of 
$h_{0,++}^{0}$ can be inferred by comparing the result from the single-channel 
analysis at $q_2^2=0$ with that from GMM's coupled-channel analysis of 
$\gamma\gamma\to\pi\pi$~\cite{GarciaMartin:2010cw}.  As shown in Figure 
\ref{fig:h00 single coupled}, the real parts begin to differ for 
$\sqrt{s}\gtrsim 0.8$ GeV, while the imaginary parts differ for 
$\sqrt{s} \gtrsim 0.5$ GeV.  The reason%
  \footnote{B.~Moussallam, private communication.} 
why the imaginary part differs at relatively small energies is because it is  
related to the real part via Watson's theorem
\begin{equation}
  \mathrm{Im}\, h_{0,++}^0(s) = 
  \pm \mathrm{Re}\, h_{0,++}^0(s) \times\tan \delta_0^0(s)\,.
\end{equation}
Near $\sqrt{s}=0.8$, the phase is close to $\pi/2$, so small variations in the 
zero of $\mathrm{Re}\, h_{0,++}^0$ can lead to a large variation in 
$\mathrm{Im}\, h_{0,++}^0$.  From a conservative viewpoint, this suggests that 
the cutoff be fixed to $\Lambda \simeq 0.8$ GeV.  However, we have checked that 
increasing the cutoff to $\Lambda = 1.2$ GeV does not lead to a difference of 
more than $\approx 7\%$ in the resulting predictions for $A_{\gamma\gamma^*}$.  
Note that this $7\%$ is the effect of a $100\%$ uncertainty on our input between 
$0.8$ and $1.2$ GeV. Since this small change is covered by our estimate of the 
systematic uncertainty, we take the larger cutoff as a benchmark value in our 
numerics and stress that only a coupled-channel analysis for this process would 
allow one to better assess this source of uncertainty and push the cutoff to yet 
higher energies. As noted in Section \ref{sec:Ktogg}, however, there are 
non-trivial difficulties in performing a coupled-channel analysis for two-body 
$K$ decays.

\begin{figure}[b]
  \centering\includegraphics[scale=0.65]{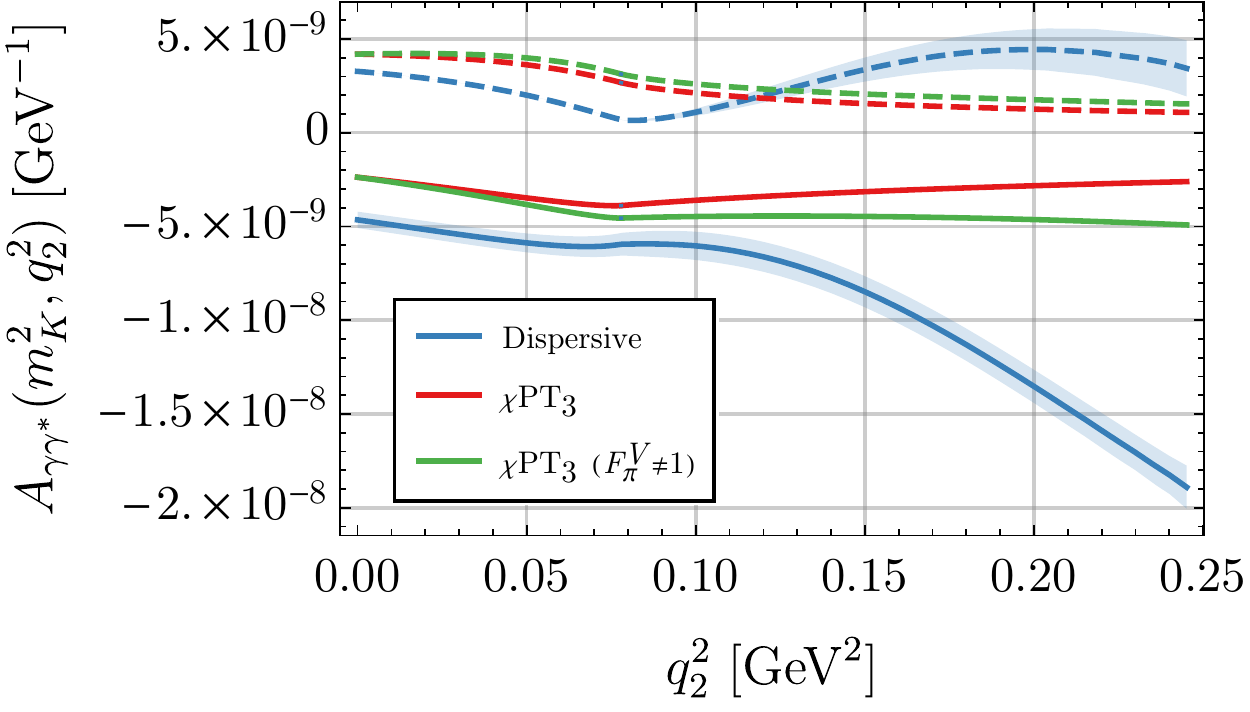}
  \caption{Dependence of the $K_S\to\gamma\gamma^*$ amplitude on the photon 
  momentum $q_2^2$ for fixed $s=\mk^2$.  The real parts are denoted by the solid 
  curves, while the imaginary parts are dashed.  The bands on the dispersive 
  results correspond to the systematic uncertainty.} \label{fig:Aggstar qsq}
\end{figure}

By combining (\ref{Aggstar born}) and (\ref{Aggstar scatt}), we obtain the 
desired result for the total $K_S\to\gamma\gamma^*$ amplitude:
\begin{equation}
  A_{\gamma\gamma^*}(s,q_2^2) = a_{\gamma\gamma^*}(q_2^2) 
  + A_{\gamma\gamma^*}^\mathrm{Born}(s,q_2^2) 
  + A_{\gamma\gamma^*}^\mathrm{scatt}(s,q_2^2)\,.
  \label{Aggstar tot}
\end{equation}
For fixed values of $q_2^2$, we first compare the predictions arising from 
(\ref{Aggstar tot}) against those of $\chi$PT$_3$. In Figure \ref{fig:Aggstar}, 
we show the energy dependence of the amplitude for three values of $q_2^2$.  As 
shown in the Figure, when $q_2^2 < 4\mpi^2$, the effect of FSI resembles that 
previously seen in $K_S\to\gamma\gamma$ (Figure~\ref{fig:Re_Im_Agg}), with the 
real (imaginary) parts enhanced (suppressed) relative to $\chi$PT$_3$.  However, 
as $q_2^2$ increases above the $\pi\pi$ threshold, the pion form factor 
$F_\pi^V$ becomes progressively more important, and both real and imaginary 
parts in the dispersive amplitude are enhanced relative to LO $\chi$PT$_3$.  
This feature can be clearly seen in Figure~\ref{fig:Aggstar qsq}, where we keep 
$s=\mk^2$ fixed and vary $q_2^2$ within the physical region
\begin{equation}
  4m_\ell^2 \leq q_2^2 \leq \mk^2
\end{equation}
of the three-body decay.
The effect of including the pion form factor in the $\chi$PT$_3$ amplitude shows 
a moderate enhancement at large $q_2^2$, especially for the real part. We also 
note that even for small values of $q_2^2$, the dispersive amplitude differs 
from $\chi$PT$_3$ due to the effects of FSI.

\begin{figure}[t]
  \centering\includegraphics[scale=0.65]{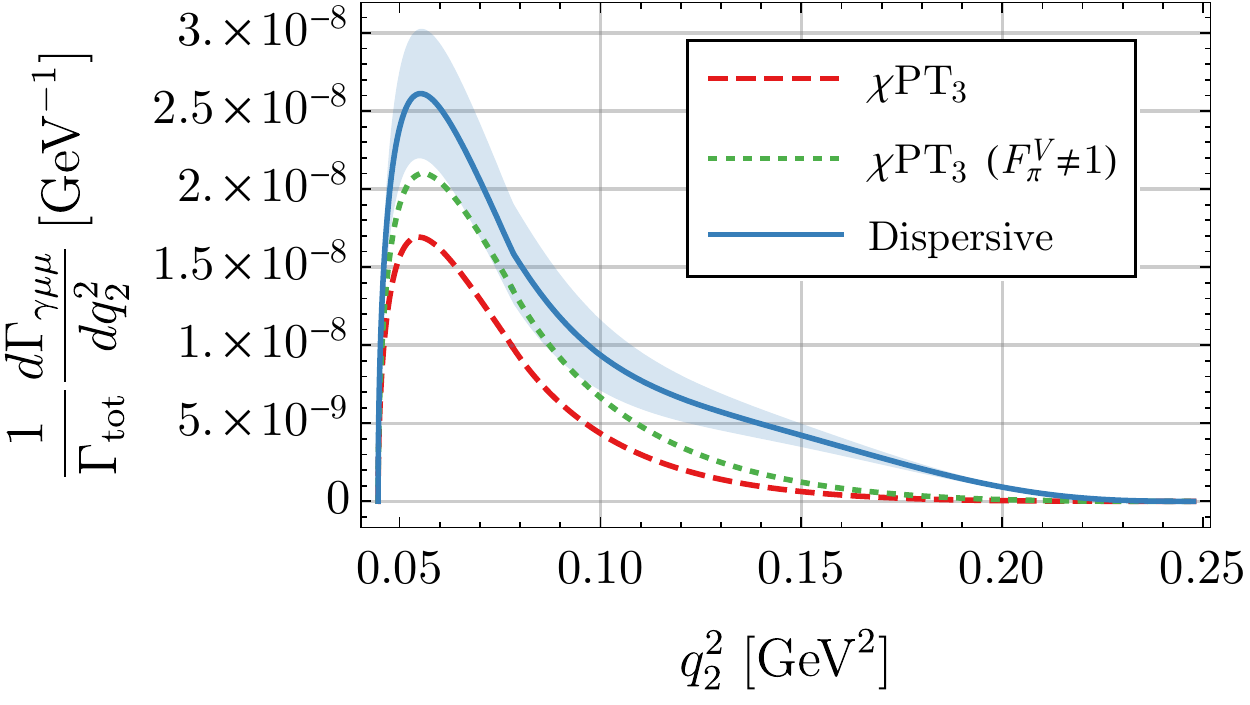}
  \caption{Differential decay width for $K_S \to \gamma\mu^+\mu^-$, normalised 
  to the total $K_S$ rate. Colour coding as in Figure \ref{fig:Aggstar qsq}.} 
  \label{fig:dGdqsq}
\end{figure}

We now consider the predictions for the $K_S \to \gamma\ell^+\ell^-$ 
decay rates.  Here the differential decay rate is~\cite{Ecker:1987hd}
\begin{equation}
  \frac{d\Gamma_{\gamma\ell\ell}}{dq_2^2} = \frac{m_K^3}{32\pi q_2^2}
  \bigg(1-\frac{q_2^2}{\mk^2}\bigg)^3\big|A_{\gamma\gamma^*}(m_K^2,q_2^2)\big|^2 
  \frac{1}{\pi} \Pi(q_2^2)\,,
\end{equation}
where the electromagnetic spectral function is given by
\begin{equation}
  \frac{1}{\pi}\Pi(q_2^2) = 
  \frac{\alpha}{3\pi} \bigg(1 + 2\frac{m_\ell^2}{q_2^2} \bigg) 
  \sqrt{1-4m_\ell^2/q_2^2}
  \,\theta(q_2^2-4m_\ell^2)\,.
\end{equation}
In Figure \ref{fig:dGdqsq}, we compare the $\chi$PT$_3$ 
prediction~\cite{Ecker:1987hd} for the differential decay rate involving muons 
against our dispersive result.  Evidently, the corrections are large for 
$q_2^2 \gtrsim 0.05$: again, this can be inferred from the $q_2^2$ behaviour 
shown in Figure \ref{fig:Aggstar qsq}.  We also see that, for this mode, the 
dominant source of the enhancement is due to the pion form factor.

The integrated rates (normalised to the total $K_S$ decay width) are shown in 
Table \ref{table:partial widths}, where the uncertainties are determined as in 
Sec.\ \ref{sec:Ktogg}, except for the subtraction constant: here we keep the 
subtraction point fixed and vary the $\chi$PT$_3$ amplitude by 30\%.  In both 
cases, the corrections are sizeable: for the electron mode we see a shift of 
$O(50\%)$, while in the muon mode we have a shift of $O(100\%)$.  The origin of 
these shifts are different in each case.  For the electron mode, the phase space 
is peaked near the origin $q_2^2=0$, so the role of $F_\pi^V$ is suppressed and 
the dominant effect is due to FSI.  On the other hand, the enhancement in the 
muon mode is predominantly due to the form factor (Figure \ref{fig:dGdqsq}).

\begin{table}[h]
\centering
\begin{tabular}{l c c } 
\hline\hline\bigstrut 
Input & $\mathrm{BR}(K_S\to\gamma e^+e^-)$ & $\mathrm{BR}(K_S\to\gamma \mu^+\mu^-)$  \\ 
\hline\bigstrut
$\chi$PT$_3$ & $3.09\times 10^{-8}$ & $7.25\times 10^{-10}$  \\
$\chi$PT$_3 \; (F_\pi^V\neq 1)$ & $3.17\times 10^{-8}$ & $9.97\times 10^{-10}$  \\
This work & $(4.38 \pm 0.57) \times 10^{-8}$ & $(1.45 \pm 0.27) \times 10^{-9}$\bigstrut  \\
\hline\hline
\end{tabular}
\caption{Predictions for the branching ratio of $K_S\to\gamma\ell^+\ell^-$.  
The second row indicates the effect of including the pion vector form 
factor $F_\pi^V$ in the $\chi$PT$_3$ amplitude.}
\label{table:partial widths}
\end{table}

%%%%%%%%%%%%%%%%%%%%
\section{Summary}%%%
\label{sec:summary}%
%%%%%%%%%%%%%%%%%%%%
Current and near-future searches for rare kaon decays are reaching sensitivities 
where a better control over the long-distance contribution to the relevant 
amplitudes is needed. Chiral perturbation theory and lattice QCD are two of the 
main tools which allow a systematic calculation of these contributions, but 
getting FSI under good control in either of these approaches is challenging. 
Dispersion relations offer a different, complementary methodology to the 
previous two, which addresses specifically the treatment of FSI. If one can 
match the dispersive and the chiral representation, and solve the dispersion 
relation, one can usually obtain much better control over FSI effects. In this 
paper, we have taken a first step in this direction by introducing a dispersive 
framework for $K_S\to\gamma\gamma$ and $K_S\to\gamma\ell^+\ell^-$.

A key feature of our analysis is that by allowing the weak Hamiltonian to
carry momentum, there is no need to extrapolate the kaon mass off-shell.
Moreover, the input for the sub-amplitudes $K_S \to\pi\pi$ and
$\gamma\gamma^{(*)}\to\pi\pi$ provide a strong constraint on the dispersive
amplitude, and when expressed in terms of measurable quantities we find
relatively small uncertainties in our final predictions.  In particular,
the Born contribution to $\gamma\gamma^*\to\pi\pi$ has a negligible
uncertainty because the pion vector form factor is known to high 
precision: for this particular contribution, going off-shell in the photon
momentum does not lead to larger uncertainties.

In general, we find that the effects due to FSI provide sizeable corrections to 
the predictions from LO $\chi$PT$_3$. For $K_S\to\gamma\gamma$, these effects 
distort the amplitude such that the relative size of the real and imaginary 
parts are interchanged. That LO $\chi$PT$_3$ predicts too large an
imaginary part can be concluded on the basis of unitarity alone and by taking
as input the experimental measurements of $K_S \to \pi \pi$ and $\gamma
\gamma \to \pi \pi$ at $s=m_K^2$: LO $\chi$PT$_3$ overshoots the correct
value by 21\%. As for the real part, we need to rely on analyticity and on a
dispersive treatment of both $K_S \to \pi \pi$ as well as $\gamma \gamma
\to \pi \pi$, where the latter is also well constrained by data. The 
uncertainties involved here are larger, but still allow us to firmly conclude 
that the prediction of LO $\chi$PT$_3$ has the correct sign (negative), but
substantially underestimates the absolute value: we obtain an enhancement
of about 70\%. This feature has been observed earlier by Kambor and 
Holstein~\cite{Kambor:1993tv}, who noted that the reasonable agreement between 
the rates from LO $\chi$PT$_3$ and experiment should be not be viewed as a 
success of the effective theory, since unitarization methods produce nearly 
identical results.  Our results confirm this observation and places it on a 
stronger footing since we do not rely on off-shell extrapolations.

For $K_S\to\gamma\ell^+\ell^-$, we found that the pion vector form factor
produces an additional source of enhancement over LO $\chi$PT$_3$. Since
the form factor is well known experimentally in both the timelike and the 
spacelike region, we can evaluate this particular correction
very reliably, which is an important outcome of this analysis. Although
less pronounced in the electron mode due to phase space suppression, we
observed a particularly large increase in the rate for the muon mode. In
view of this result, we believe the muon mode has good prospects of being 
observed at the projected sensitivities of KLOE-2.

In our analysis, we have restricted ourselves to the case where at most one 
photon is off-shell. It would be interesting to extend our dispersive 
framework to the doubly off-shell amplitude $K_S \to \gamma^*\gamma^*$, which 
provides the dominant contribution to the rare decay $K_S \to\ell^+\ell^-$. For 
the muon mode, LHCb~\cite{Aaij:2012rt} has recently placed an upper bound 
on the rate $\mathrm{BR}(K_S\to\mu^+\mu^-) < 9 \times 10^{-9}$, and future 
upgrades are expected to improve the sensitivity down to 
$O(10^{-10})$~\cite{Yamanaka:2014xta}. Given that a signal well above 
$10^{-11}$ has been claimed~\cite{Isidori:2003ts} to be clear evidence of 
physics beyond the SM, determining the role of FSI in this mode will be
essential in order to draw definite conclusions regarding the SM
background. Work in this direction is currently in progress.

%%%%%%%%%%%%%%%%%%%%%%%%%%%%
\section*{Acknowledgements}%%
%%%%%%%%%%%%%%%%%%%%%%%%%%%%
We are indebted to Bachir Moussallam for providing us with numerical data and 
code from his analyses of $\gamma\gamma\to\pi\pi$ and $\gamma\gamma^*\to\pi\pi$.  
We also thank him for extensive correspondence and helpful suggestions on the 
work presented in this paper.  We thank Martin Hoferichter and Peter Stoffer for 
useful discussions and correspondence; we also thank them and Gerhard Ecker 
and Toni Pich for providing comments on the manuscript.  
We thank Bastian Kubis for insightful remarks on the role of Born 
terms in $K_S\to\gamma\gamma^*$. 
The authors are 
grateful to the Mainz Institute for Theoretical Physics (MITP) for the 
hospitality and partial support during the completion of this work.  This work 
was partially funded by the Swiss National Science Foundation.

\end{document}